\documentclass[letterpaper,twocolumn,10pt]{article}

\usepackage[linesnumbered,ruled,vlined]{algorithm2e}
\usepackage{multirow}
\usepackage[table,xcdraw]{xcolor}
\usepackage{array}
\usepackage{adjustbox}
\usepackage{graphicx}
\usepackage{amsmath}

\usepackage[T1]{fontenc}
\usepackage{usenix2019_v3}
\usepackage{newtxtext,newtxmath}

\usepackage{tikz}
\newcommand*\circled[1]{%
  \tikz[baseline=(char.base)]{
    \node[
      shape=circle,
      draw,
      inner sep=0.8pt,         % Adjust inner padding
      font=\sffamily\small,  % Optional: adjust font style and size
      align=center,
      anchor=center          % Ensure the node is centered
    ] (char) {#1};
  }%
}
\usepackage{amsmath}
\usepackage{authblk}
\usepackage{subfig} % subfloat

\usepackage{filecontents}
\usepackage{array}
\newcolumntype{C}[1]{%
  >{\centering\arraybackslash\hspace{0pt}}%
  m{#1}%
}

\begin{document}
%-------------------------------------------------------------------------------

%don't want date printed
\date{}

\title{Adaptive Migration Decision for Multi-Tenant Memory Systems}

\author{
Hyungjun Cho\textsuperscript{*} \quad
Igjae Kim\textsuperscript{*} \quad
Kwanghoon Choi\textsuperscript{*} \quad
Hongjin Kim\textsuperscript{*} \quad
Wonjae Lee\textsuperscript{\dag} \quad
Junhyeok Im\textsuperscript{\dag} \quad
Jinin So\textsuperscript{\dag} \quad
Jaehyuk Huh\textsuperscript{*} \\
\textsuperscript{*}KAIST \quad \textsuperscript{\dag}Samsung Electronics\\
whgudwns202@gmail.com, 
\{ijkim, khchoi, hjkim\}@casys.kaist.ac.kr \\
\{wj28.lee, junhyeok.im, jinin.so\}@samsung.com,  jhhuh@kaist.ac.kr
}
\maketitle

\begin{abstract}
Tiered memory systems consisting of fast small memory and slow large memory have emerged to provide high capacity memory in a cost-effective way. The effectiveness of tiered memory systems relies on how many memory accesses can be absorbed by the fast first-tier memory by page migration. The recent studies proposed several different ways of detecting hot pages and migrating them efficiently.
However, our investigation shows that page migration is not always beneficial as it has the associated cost of detecting and migrating hot pages. When an application is unfriendly to migration, it is often better not to migrate pages at all. 
Based on the observation on migration friendliness, this paper proposes a migration control framework for multi-tenant tiered memory systems. First, it proposes a detection mechanism for migration friendliness, using per-page ping-pong status. Ping-pong pages which are promoted and demoted repeatedly in a short period of time tells migration effectiveness. Based on their change behaviors, migration is stopped or continued.
After the page migration is stopped, the second mechanism detects changes of memory access patterns in a low cost way to determine whether migration needs to be resumed. Finally, as each application has a different behavior, our framework provides per-process migration control to selectively stop and start migration depending on application characteristics.
We implement the framework in the Linux kernel. The evaluation with a commercial CXL-based tiered memory system shows that it effectively controls migration in single and multi-tenant environments.

\end{abstract}

%-------------------------------------------------------------------------------
\section{Introduction}
%-------------------------------------------------------------------------------

To address ever-growing demands for memory capacity,  tiered memory systems have emerged as a cost-effective way to increase the effective memory capacity~\cite{computeexpresslink,infiniswap,fastswap}. The tiered memory systems consist of a low latency and high bandwidth first-tier memory backed by the slower second-tier memory. Recent new interconnect technologies such as CXL ~\cite{computeexpresslink,ji2024demystifying,sun2023demystifying} enabled the tiering of memory components by combining direct DRAM modules with CXL-connected expander memory. To create an illusion of fast high capacity memory with tiering, the majority of memory accesses must occur in the first tier memory effectively. To place memory pages likely to be accessed in the near future in the first tier memory, the recent system solutions rely on page migration with hot page detection~\cite{maruf2023tpp,lee2023memtis,xu2024flexmem,xiang2024nomad,song2023freqtier,bergman2022reconsidering,dulloor2016data,gupta2015heterovisor,hildebrand2020autotm,kannan2017heteroos,maruf2022multi,oh2021maphea,yan2019nimble,heo2020adaptive}.  

This study first investigates the effectiveness of page migration by the recent approaches and finds that dynamic page migration can be harmful in certain cases. Page migration mechanisms have their own costs. Profiling and hot page detection cause hint fault handling overheads. When a migration occurs, it consumes the memory and computation bandwidth to exchange pages and update page tables with TLB shootdowns. Unless page migrations can effectively increase hits on the first tier memory, the page migrations only cause performance degradation. 

Such unnecessary page migration can often occur in the current tiered memory systems. The working set of an application can exceed the capacity of the first tier memory, causing continuous page eviction and migration between the first and second tier memories. In addition, the hot detection uses per-page access history either collected by hint faults~\cite{van2014automatic} or hardware-based instruction sampling~\cite{pebs}. However, the limited information used by hot page detection cannot accurately select hot pages. We observe stopping migration often produces better performance than incurring migration costs without any benefit of migration.

However, stopping migration must be cautiously controlled to avoid blocking effective page migration. The framework requires three main mechanisms. i) It should decide whether the current page migrations for an application are beneficial or harmful. ii) As the application behavior changes after migration stops, it must be able to resume the page migration when the migration becomes useful. iii) It must consider multiple applications that can be co-running in common multi-tenant environments.

To address the limitations of recent page migration approaches, this paper proposes a migration friendliness-aware page management for tiered memory systems. The framework determines migration friendliness by assessing whether migration can be stabilized. It uses the ping-pong migration when a page is promoted and demoted repeatedly in a short period of time as the key indicator. For each page, the ping-pong migration status is tracked, and how the number of ping-pong page changes determines migration friendliness.

The second mechanism is to restart the stopped migration. The restart mechanism must detect changes in page access behaviors without using the costly hint fault mechanism. We use sampled access-bit information in page tables to find whether page access patterns change significantly.

Finally, the migration control mechanism must be aware of multi-tenancy. Each application has a different memory access behavior, and thus stopping migration globally can harm migration-friendly applications. Our framework selectively tracks migration friendliness for processes and determines whether migration must be turned on or off for each process. Unlike our work, the prior hotness-based page migration globally determines hot pages without considering the different behaviors of co-running applications.

We have implemented the migration friendliness-aware management framework in Linux v5.15. We use a commercial CXL-based tiered memory system for our evaluation.
For single tenant scenarios, when an application is migration-friendly, it provides a comparable performance with the best of the recent page migration approaches. If an application is migration-unfriendly, it effectively turns off migration, providing an average of 14.8\% performance improvement compared to NOMAD. And if an application is migration-friendly, our scheme provides an average of 36.0\% improvement.
For multi-tenant scenarios, our scheme achieves up to 72.0\% performance improvement compared to NOMAD, which demonstrates that migration toggling is effectively applied at the per-process level considering their migration friendliness.

To complement recent studies to support hotness-based page migration, this paper advocates the need for controlling page migration for tiered memory systems. The source code will be publicly available after publication. The main contributions of the paper are as follows.

\begin{itemize}
\item The paper shows that page migration is not always beneficial as the current mechanism has non-negligible costs for detecting and migrating hot pages. 
\item The paper proposes a low cost detection mechanism for deciding the migration friendliness of an application.
\item The paper proposes a migration restart mechanism by tracing the changes in memory access behavior.
\item The paper advocates the need for controlling page migration differently for each application in multi-tenant environments.
\end{itemize}

%-------------------------------------------------------------------------------
\section{Background}
%-------------------------------------------------------------------------------
% \input{figure/figure}
% \input{table/table}
\subsection{Tiered Memory System}
%-----------------------------------

As applications requiring large amounts of memory, such as large language models (LLMs) and high performance computing (HPC), continue to proliferate~\cite{ousterhout2010case,Butler2012MemoryShortage,touvron2023llama,openai2023chatgpt}, the importance of system memory capacity has grown significantly. 
However, the maximum memory capacity a system can support is inherently limited by physical constraints, which restrict the execution of such large-scale applications. 

To address such challenges, tiered memory systems have been widely studied.
These systems comprise multiple memory types with different characteristics, typically featuring high-tier memory that offers low latency and limited capacity, alongside low-tier memory that provides higher latency and larger capacity~\cite{sun2023demystifying,ji2024demystifying}.
For example, Compute Express Link (CXL)~\cite{computeexpresslink} memory expands system memory via PCIe lanes with cache-line granularity.
In addition, heterogeneous memory types, such as High-Bandwidth Memory (HBM)~\cite{jun2017hbm} and Optane memory~\cite{izraelevitz2019basic}, are integrated with CPUs to establish memory tiers within a single system.

\subsection{Page Hotness Detection}

% \begin{profmethodsummary}
% \end{profmethodsummary}
\begin{table}[t]
    \centering
    \resizebox{\columnwidth}{!}{
        \begin{tabular}{c|c|c|c|c|c}
            \hline
            \textbf{} & \textbf{Linux~\cite{bharata2024patch}} & \textbf{TPP~\cite{maruf2023tpp}} & \textbf{NOMAD~\cite{xiang2024nomad}} & \textbf{MEMTIS~\cite{lee2023memtis}}  & \textbf{Ours} \\ \hline
            Transparent Profiling  & O & O & O & $\triangle$ & O \\ \hline
            Multi-tenant & $\triangle$ & $\triangle$ & $\triangle$ & $\triangle$ & O \\ \hline
            Conditional Migration & X & X & X & X & O \\ \hline
        \end{tabular}
    }
    \caption{Compared to prior works, our approach enables per-process profiling for multi-tenant scenario and introduces conditional migration to mitigate detrimental migration.}
    \label{tab:profmethodsummary}
\end{table}

In tiered memory systems, strategically allocating data across different memory tiers to leverage each tier's unique advantages—known as page migration—is essential.
Numerous recent research~\cite{maruf2023tpp,lee2023memtis,xu2024flexmem,xiang2024nomad,song2023freqtier,bergman2022reconsidering,dulloor2016data,gupta2015heterovisor,hildebrand2020autotm,kannan2017heteroos,maruf2022multi,oh2021maphea,yan2019nimble,heo2020adaptive} target tiered memory systems, typically employing three components for page migration: profiling mechanisms, migration policies and migration mechanisms.

\noindent
\textbf{Page access profiling:}
Identifying frequently accessed pages is crucial for migrating them to their appropriate locations.
However, it is challenging to detect their frequency because memory accesses are handled transparently by the hardware memory management unit (MMU) and are not directly visible to the software. 
To address this, prior research on tiered memory systems has commonly adopted two methodologies: hint fault and PEBS~\cite{pebs}.

Hint fault serves as the foundational mechanism of AutoNUMA~\cite{van2014automatic} which is the default memory migration solution in the Linux kernel. 
It operates entirely in software by periodically setting a bit on the page table entries (PTEs) of a specific amount of memory. 
If access occurs on such PTEs, a minor page fault, referred to as a hint fault, is triggered.
TPP~\cite{maruf2023tpp} and NOMAD~\cite{xiang2024nomad} leverage modified hint fault to detect page access information.
In contrast, PEBS (Precise Event-Based Sampling) utilizes dedicated hardware for access monitoring. 
PEBS captures predefined hardware events (e.g., \texttt{LLC\_LOAD\_MISS}, \texttt{STLB\_STORE\_MISS}) and records them in hardware-managed buffers. 
It defines the sampling period; higher sampling frequencies generate more samples but also increase overhead.
MEMTIS~\cite{lee2023memtis} utilizes PEBS to profile memory access patterns.

\noindent
\textbf{Migration policy by hotness detection:}
A migration policy specifies whether individual pages should be relocated to a different memory tier based on collected memory access information. The principal objective of page migration is to place frequently accessed (“hot”) pages in faster memory (commonly DRAM). Pages that are accessed more often are considered hot, whereas those accessed less frequently are treated as cold.

A commonly employed technique for identifying hot pages involves tracking access counts. For instance, MEMTIS maintains a per-page access count and constructs a histogram from these counts, selecting the top-\texttt{N} entries as hot pages according to the capacity of the fastest memory tier. 
In contrast, TPP and NOMAD leverage the Linux default \texttt{lruvec} and categorize pages in the active list as hot. 
However, page migration between active list and inactive list occurs under memory pressure—such as page reclamation or \texttt{kswapd} operations which access the access bit written in page table entry by MMU —meaning that TPP’s hot-page selection still relies heavily on per-page access counts.

\begin{figure}[t!]
    \begin{center}
    \includegraphics[width=0.3\textwidth]{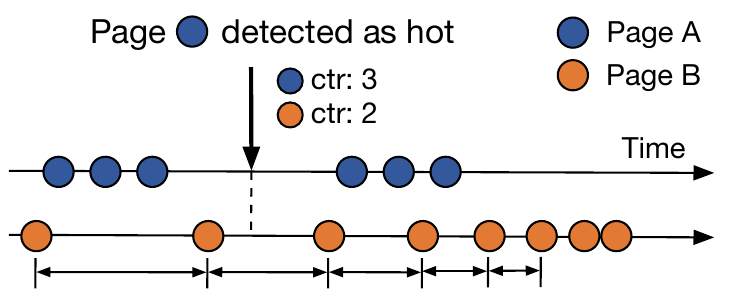}
    \end{center}
    \vspace{-0.2in}
    \caption{The limitation of counter-based page hotness detection.}
    \label{fig:accesscount}
    \end{figure}

Furthermore, this access counter-based policy does not always hold. 
For instance, Figure~\ref{fig:accesscount} illustrates the access patterns of two pages in a system where only one page can be promoted to the fast tier. 
Although page A has a higher access count at the time of selection and is therefore promoted, a broader view of the access patterns reveals that the access intervals for page B consistently decrease, indicating that page B would be a more critical candidate for promotion. This example highlights a key limitation of access count-based approaches, which are commonly employed in tiered memory systems. MEMTIS~\cite{lee2023memtis} addresses this issue by periodically "cooling" the access count histogram to adjust the hot set dynamically. However, if the cooling period is too short, the system may fail to capture the overall access pattern, while an overly long cooling period can still lead to suboptimal decisions. 

To overcome the limitation of counter-based policy, \textit{Memory tiering} in Linux employs the notion of a “hint fault latency”~\cite{bharata2024patch} which considers temporal access pattern. 
Hint fault latency is defined as the time difference between the time a page is poisoned to trigger a hint fault (using \texttt{PROT\_NONE}) and the actual access to that page.
The page is promoted only when the hint fault latency is shorter than the pre-defined static threshold, thereby considering its temporal information.

% \begin{accesscountfigure}
% \end{accesscountfigure}

\subsection{Page Migration Mechanism}
\noindent
\textbf{Migration mechanisms:}
The migration mechanism indicates the method by which memory pages are transferred between different memory tiers. In Linux, page migration involves several distinct steps. Initially, the page table entry (PTE) corresponding to the target page must be unmapped from the page table. Following the unmapping, a Translation Lookaside Buffer (TLB) shootdown is initiated via inter-processor interrupts to ensure application consistency. 
Once the contents of the page are copied, the PTE is remapped to the new location. During the migration process, the application is unable to access the corresponding page.
In both Linux and TPP, pages are migrated when the application accesses incur a minor-faulted page. 
At this point, the kernel handles the hint fault, and if the page is determined to be a candidate for promotion to a faster tier, it is migrated directly while blocking the application until migration completes.

NOMAD~\cite{xiang2024nomad} mitigates the overhead from synchronous migration by decoupling the migration process from the application’s critical path. 
It allows the application to continue accessing the original page in the slower tier during promotion, thereby allowing migration to proceed asynchronously. 
If the migrated page becomes dirty after promotion, the updated data is copied from the slower tier to ensure consistency.
MEMTIS~\cite{lee2023memtis} employs two additional kernel threads, one for promotion and another for demotion to migrate pages asynchronously in the background, minimizing the impact on application performance.  

\noindent
\textbf{Page demotion:}
In hotness-based page migration, frequently accessed pages are continually promoted to higher-tier memory. As this process continues, page demotion is required to reserve additional space in the higher-tier memory for new allocations. In Linux, each node utilizes a concept known as “watermarks” to prevent out-of-memory (OOM) situations. Once the amount of allocated memory surpasses the watermark, the kernel invokes the \texttt{kswapd} daemon, which either reclaims pages or demotes them. Specifically, \texttt{Kswapd} selects tail pages from the LRU-based inactive list located in higher-tier memory and either demotes them to lower-tier memory or swaps them out to storage, thereby freeing up additional space in the higher-tier memory.

Based on such mechanisms, TPP~\cite{maruf2023tpp} and NOMAD~\cite{xiang2024nomad} introduce an additional watermark beyond the conventional threshold. 
This extra watermark triggers kswapd earlier—before the memory usage reaches the default watermark—thereby maintaining sufficient free space in the higher-tier memory to accommodate future promotions. 
The approach of MEMTIS diverges from the aforementioned method and instead aligns more closely with its own page promotion strategy. 
Specifically, MEMTIS consults a histogram compiled during the promotion phase to guide demotion: it demotes higher-tier pages that do not surpass a specified access-frequency threshold which is determined based on higher-tier memory capacity.

%-------------------------------------------------------------------------------
\section{Motivation}
\label{sec:figs}
%-------------------------------------------------------------------------------

\subsection{Migration-friendliness}

\begin{figure}[t]
    \centering
    \includegraphics[width=0.5\textwidth]{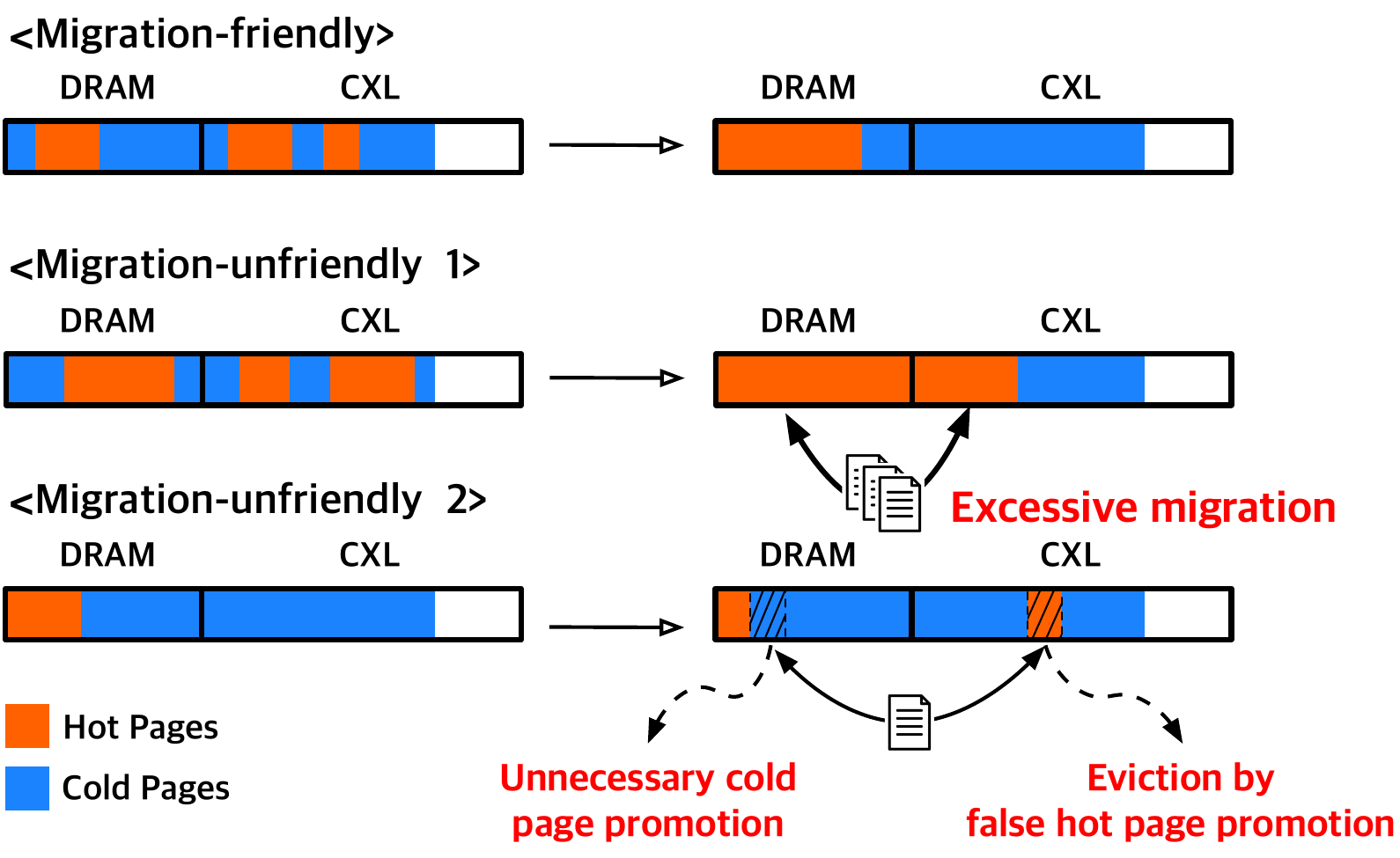}
    \caption{Migration-friendly/unfriendly cases.}
    \label{fig:migrationfriendliness}
\end{figure}

Many studies suggest that page migration can mitigate the performance degradation associated with the access latency of slower memory tiers by relocating frequently accessed pages to faster memory.  However, it is crucial to consider additional factors to gain performance improvement by page migration. Figure~\ref{fig:migrationfriendliness} illustrates scenarios in which page migration can be advantageous or detrimental.

Page migration requires two conditions to enhance application performance. First, the set of hot pages must be clearly defined. A hot page set comprises pages that exhibit a significantly higher access ratio compared to other pages. Second, the hot page set size must be constrained to fit within the capacity of the fast-tier memory.

As illustrated in the upper section of figure~\ref{fig:migrationfriendliness}, referred to as a migration-friendly case, the entire hot page set resides in DRAM after sufficient migration. This placement ensures that most accesses occur within the low-latency DRAM, thereby enhancing performance. However, the situation changes in cases that are migration-unfriendly. Firstly, if the hot set exceeds the capacity of DRAM, it will continuously attempt to relocate the remaining hot pages from CXL to DRAM, resulting in additional overhead associated with the migration process.
MEMTIS~\cite{lee2023memtis} addresses this issue by limiting the size of the hot page set to the available capacity of DRAM within its policy framework. 
Nevertheless, it remains uncertain whether the pages identified as hot by MEMTIS truly align with the optimal set of hot pages from an ideal perspective. 
Secondly, if an application exhibits a random access pattern, most portions of WSS will be uniformly accessed.
In this case, pages with similar hotness levels are promoted to DRAM, but they may not be accessed frequently enough even with MEMTIS policy.
Consequently, these pages may be demoted back to the slower memory tier shortly after promotion, leading to repetitive cycles of promotion and demotion.

As previously discussed, the intrinsic characteristics of an application significantly influence its migration-friendliness. 
However, this does not universally apply across all hardware configurations. 
Variations in system attributes, such as DRAM size, bandwidth, and access latency to the slower memory tier, can cause an application that appears migration-unfriendly on one system to appear migration-friendly on another.

% \begin{motiv2figure}
% \end{motiv2figure}

\begin{figure}[t]
    \centering
    \subfloat[Varying results of benchmarks with different migration schemes.]{
        \includegraphics[width=1.0\linewidth]{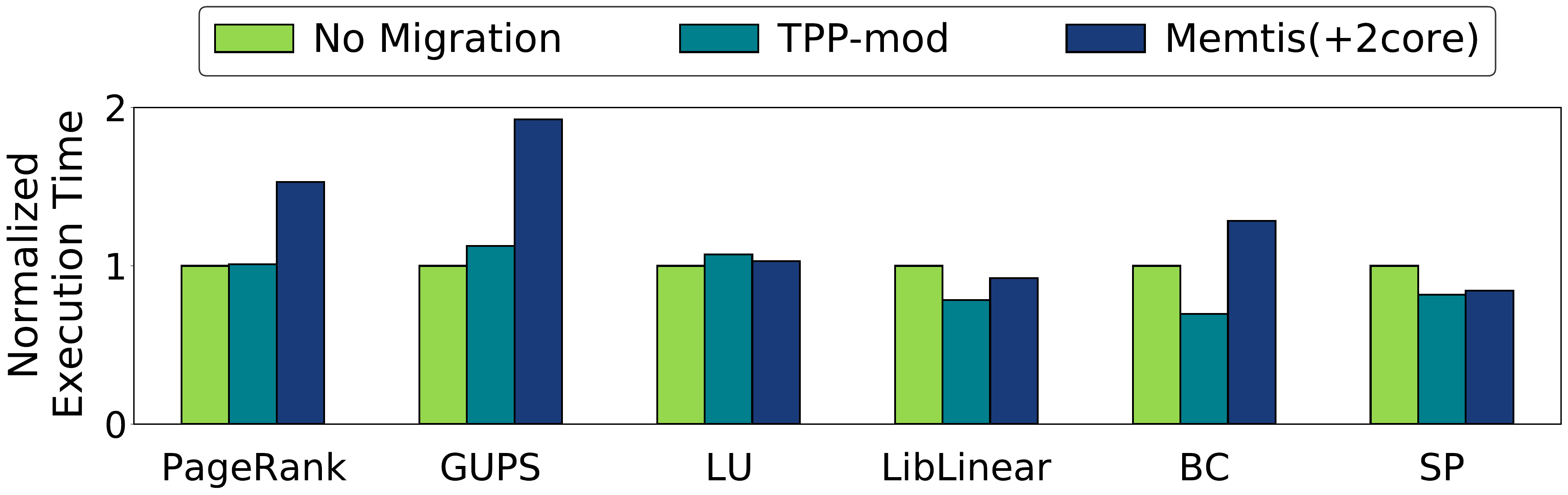}
        \label{fig:motiv_2_1}
    }
    \vspace{-0.1in}
    \hfill
    \subfloat[Execution time of GUPS and LU with increasing DRAM capacity.]{
        \includegraphics[width=1.0\linewidth]{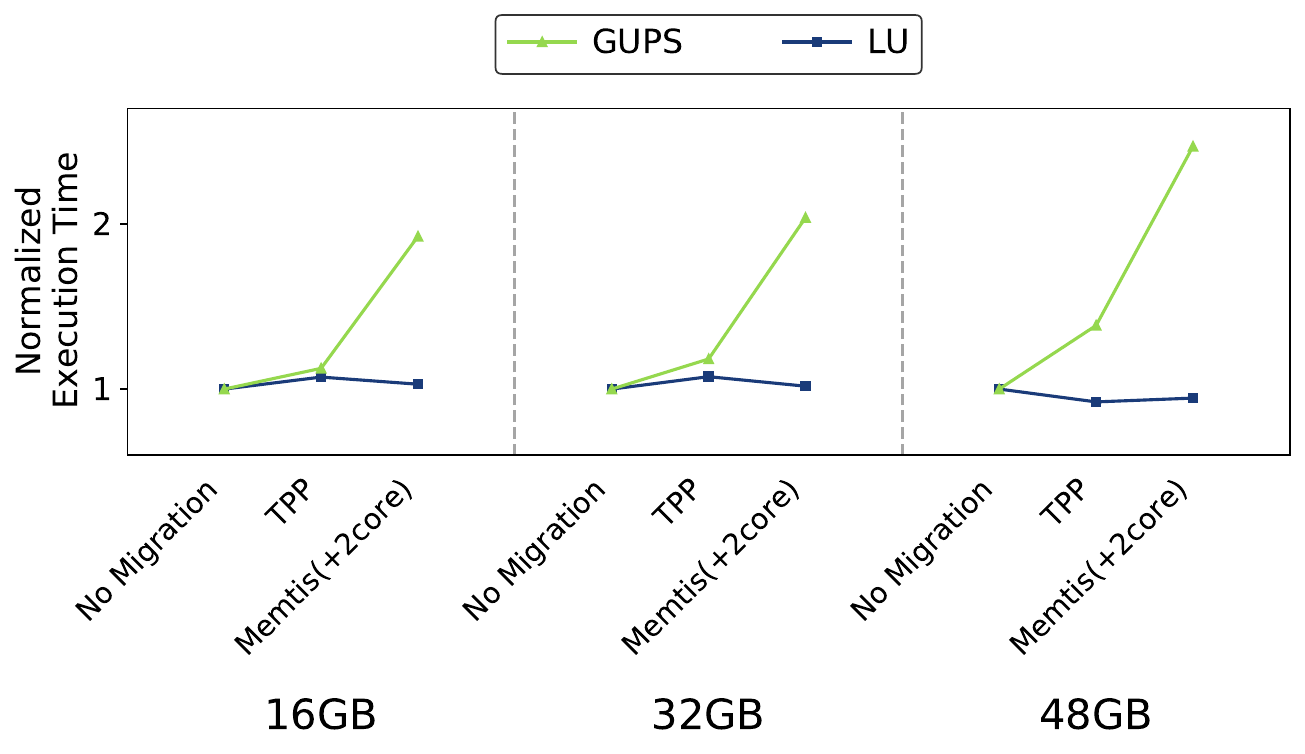}
        \label{fig:motiv_2_2}
    }
    \caption{The evaluation results demonstrate the existence of migration-friendliness. Performances are normalized to No Migration.}
        \label{fig:motiv_2}
\end{figure}
\noindent
\textbf{Benchmarks evaluation:}
Figure~\ref{fig:motiv_2_1} shows the results of No Migration, Modified TPP (TPP-mod, discussed in Section~\ref{migrationopt}), and MEMTIS across the benchmarks.  
It shows that performance depends on both their migration-friendliness and the efficiency of the migration scheme employed. 
Specifically, in the cases of GUPS and LU, changing the DRAM capacity results in distinct performance behaviors. Figure~\ref{fig:motiv_2_2} shows the normalized execution time of GUPS and LU with varying DRAM size. 
In the case of GUPS, despite having a total RSS of 60GB, increasing the DRAM capacity up to 48GB does not produce any performance improvement. 
This result suggests the absence of a true hot page set, which demonstrates that migration is ineffective across most hardware configurations for GUPS. 
Conversely, for LU, performance degrades with DRAM capacities of 16GB and 32GB but improves when the DRAM capacity is increased to 48GB. 
This scenario likely involves a distinguishable hot page set whose size exceeds 32GB but remains below 48GB, indicating that migration-friendliness depends on the DRAM capacity in such case.
These results demonstrate that the migration-friendliness of workloads can change based on system configurations.
Furthermore, page migration for migration-unfriendly workloads can result in performance degradation.
 
\subsection{Migration Overhead of Unfriendly Cases}
There are two primary factors by which migration can cause significant performance degradation under migration-unfriendly cases.
First, it can block the application control flow, which occurs during the migration process itself. 
Second, overheads from inaccurate hot page detection can arise. To measure these factors in detail, we conduct an experiment using a single-threaded microbenchmark where the WSS exceeds twice the DRAM capacity. This experiment is based on an Optane-based tiered memory system where the difference in access latency is 100ns, and TPP-mod is employed as a migration scheme. 

\textbf{Overhead of migration procedure:} 
The results show that fault handling without migration under TPP's policy consumed 4\textasciitilde5µs, whereas fault handling with migration consumed 13\textasciitilde28µs. According to TPP's criteria, if migration occurs after two hint faults, a fault handling time of 17,000–33,000ns is required. 
The entire migration process within fault handling takes 9\textasciitilde23µs, during which the access is blocked, waiting for the migration to complete. 
Further measurements of the main migration steps—allocation, unmapping, copying, and remapping-indicate that each step consumes approximately 1\textasciitilde2µs, 2\textasciitilde4µs, 5\textasciitilde7µs, and 2\textasciitilde3µs, respectively. 
Although the unmap process is generally recognized as the most resource-intensive task due to triggering TLB shootdowns, the limited bandwidth results in significantly higher time consumption for write operations, leading to relatively longer durations during the copy phase.

\textbf{Migration-friendly/unfriendly cases:}
Under the above conditions, a comparison is made between scenarios where migration promotes true hot pages and those where it promotes false hot pages. When a single page is selected and promoted as a hot page, one of the existing pages in DRAM must be demoted to create free space. 
In this case, there are two possibilities: the demoted page remains a true hot page or turns out to be a true cold page. 
If a true hot page is promoted and the demoted page also becomes a true cold page, it is the desired case as this alignment maximizes the effectiveness of the migration.

However, if the demoted page remains a true hot page (figure \ref{fig:migrationfriendliness}, migration-unfriendly 1), the overhead caused by blocking migration steps is highlighted. Depending on the microbenchmark settings, the hot page is accessed A times, resulting in a time gain of 100ns × A for access to the promoted page and a time loss of 100ns × A for access to the demoted page. These gains and losses compensate for each other, leaving only the migration overhead of 17,000-33,000ns from fault handling and 9,000-14,000ns from demotion as a net loss. In other words, even if the true hot page is correctly identified and all pages already in DRAM are true hot pages, only the migration overhead from demotion and promotion remains a loss.

Furthermore, if migration promotes a false hot page by demoting one of the pages in DRAM (figure \ref{fig:migrationfriendliness}, migration-unfriendly 2), a different outcome arises. If a cold page is demoted, it just leaves fault handling overhead from promotion and migration overhead from demotion as before. However, if a hot page is demoted, the total overhead includes the two fault handling times required for promotion, the migration overhead from demotion, and an additional time loss proportional to 100ns × A due to losing the hot page in DRAM.

Even if both promotion and demotion are executed in the background as implemented in MEMTIS, or if the migration sequence is overlapped and the costs associated with demotion are partially reduced through shadowing as in NOMAD, the time consumed by page migration remains non-negligible from an overall perspective if migration does not make true hot page promotion and true cold page demotion.

Furthermore, it is challenging to determine whether a workload is migration-friendly on a target system. Therefore, if migration-friendliness can be assessed cost-effectively at runtime and all migration-related features are disabled when a workload is identified as migration-unfriendly, we can anticipate that performance can be improved even with migration-unfriendly workloads.

\subsection{Multi-tenant Environment}
%-----------------------------------

\begin{figure}[t]
    \centering
    \subfloat[Effect of page migration on multi-tenancy with varying execution start time.]{
        \includegraphics[width=1.0\linewidth]{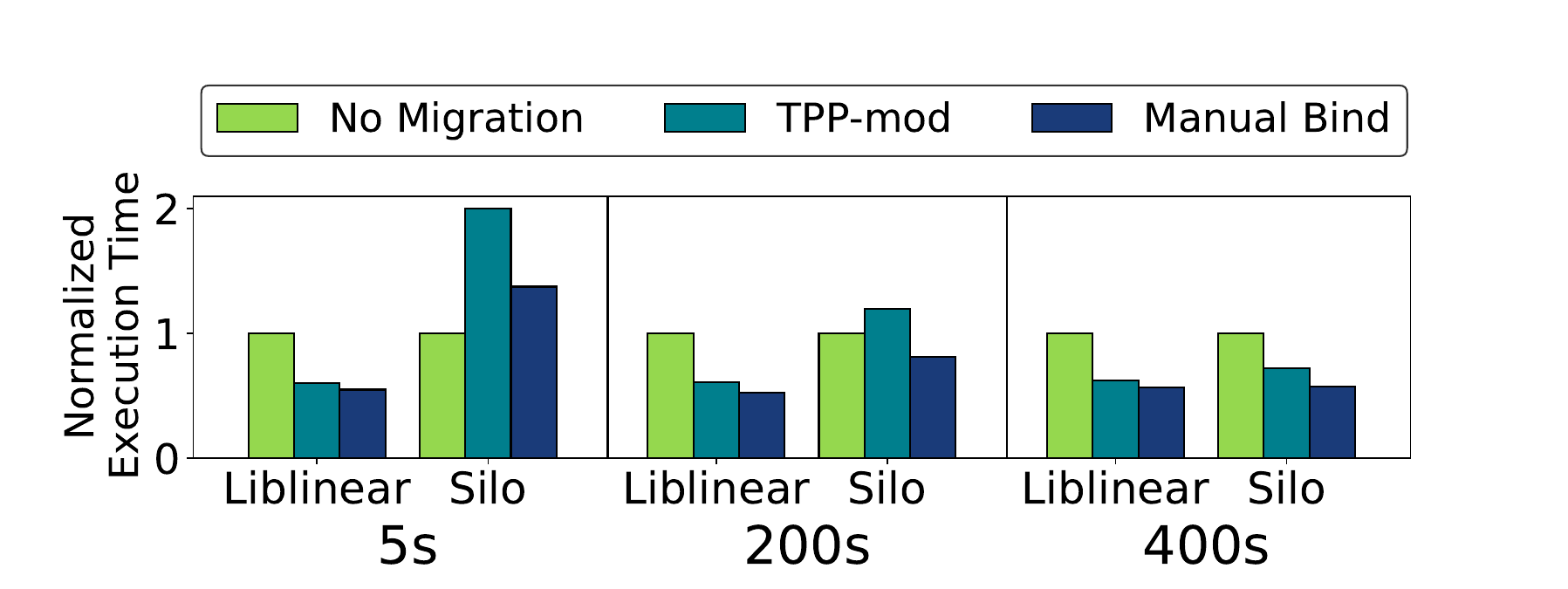}
        \label{fig:motiv_3_1}
    }
    \vspace{-0.1in}
    \hfill
    \subfloat[Single-tenancy evaluation. It shows Liblinear is migration-friendly, and Silo is migration-unfriendly.]{
        \includegraphics[width=0.9\linewidth]{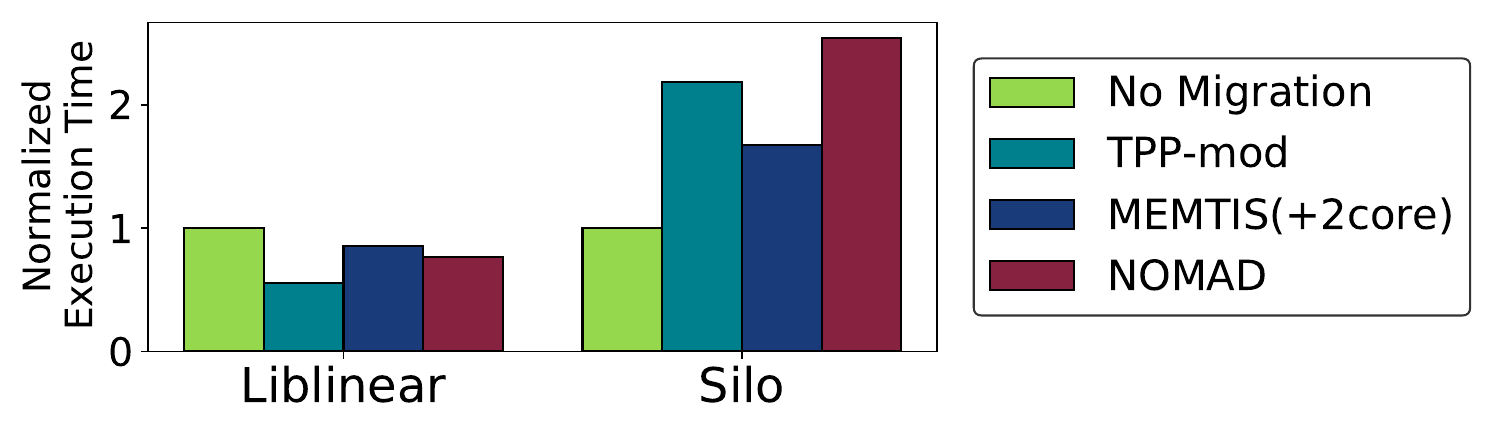}
        \label{fig:motiv_3_2}
    }
    \caption{Page Migration on multi-tenancy. Performances are normalized to No Migration.}
        \label{fig:motiv_3}
\end{figure}

If applications are executed in a multi-tenant system, the interfered access pattern of multiple applications prevents the system from identifying each application's migration-friendliness, even if each application is aware of its single-tenancy migration-friendliness based on its respective hardware configuration.
For example, although TPP is designed to support multi-tenancy, it has inherent limitations because the lruvec that TPP leverages to maintain access to information is managed globally rather than on a per-application basis. 
When migration-friendly and migration-unfriendly workloads are executed simultaneously, hot pages from friendly workloads may be demoted to the inactive list by frequently accessed pages from unfriendly workloads, even if those pages are not a true hot set in its process. 
Furthermore, the promotion of these unfriendly pages may evict a true hot set of migration-friendly workloads located within DRAM. 
Similarly, in the case of MEMTIS, where migration-friendly and migration-unfriendly workloads are executed simultaneously, and the pages from unfriendly workloads dominate in terms of access counts, the top-n entries in the histogram may become entirely occupied by unfriendly pages. 
This scenario poses a significant challenge to achieve optimal performance.

To validate the aforementioned scenario, an experiment is conducted by simultaneously running a friendly workload (Liblinear:15 threads) and an unfriendly workload (Silo-single thread) which is represented in Figure~\ref{fig:motiv_3_2}. 
Execution times are measured under three conditions: No Migration, TPP-mod, and Manual Bind (migration is applied on Liblinear while Silo is bound to a CXL node). 
The longer-running Liblinear was initiated first, and in cases 1, 2, and 3, FT was started after 5 seconds, 200 seconds, and 400 seconds, respectively.

Figure~\ref{fig:motiv_3_1} shows results for multi-tenant applications. The experimental results show that manual bind makes the best performance for all scenarios. In the case of the migration-friendly workload Liblinear, TPP consistently improves performance. However, for Silo, the best performance is observed when the execution start time is delayed, allowing Liblinear to occupy the entire DRAM at its initial start.
 We analyze that, even though TPP manages hot pages through a global \textit{lruvec}, the number of access on Liblinear’s hot page set is sufficient to suppress hot page determination of Silo. These results demonstrate that, in a multi-tenant environment, it is necessary to decide whether to perform migration independently for each process based on the individual migration-friendliness of the workloads, rather than using a global policy to determine migration execution.

% _
% \input{algorithm/algorithm}
%-------------------------------------------------------------------------------
\section{Design}
\label{sec:figs}
%-------------------------------------------------------------------------------

\subsection{Overview}
Our design dynamically toggles the migration based on migration-friendliness in per process manner.
The design of our code uses TPP as a baseline.  Also, we made a small patch on TPP named 'Modified second chance LRU' that bypasses \texttt{pagevec}, where the page waits to move to the active list, to prevent excessive fault handling(TPP-mod). 
Based on TPP-mod, we add a dynamic migration toggle feature that stops the migration based on runtime-evaluated migration friendliness and restarts the migration by identifying variations of access pattern. Additionally, multi-tenancy environments are efficiently supported by incorporating per-process considerations.

\subsection{Migration Stop}
%-----------------------------------
To assess the migration-friendliness of a workload at runtime, a low-cost metric is essential. After evaluating various metrics, the optimal indicator is the per-page ping-pong rate. It refers to the scenario where specific pages repeatedly undergo promotion and demotion between the DRAM and CXL memory.

\begin{figure}[t]
    \centering
    \subfloat[Delta]{%
        \includegraphics[width=0.48\linewidth]{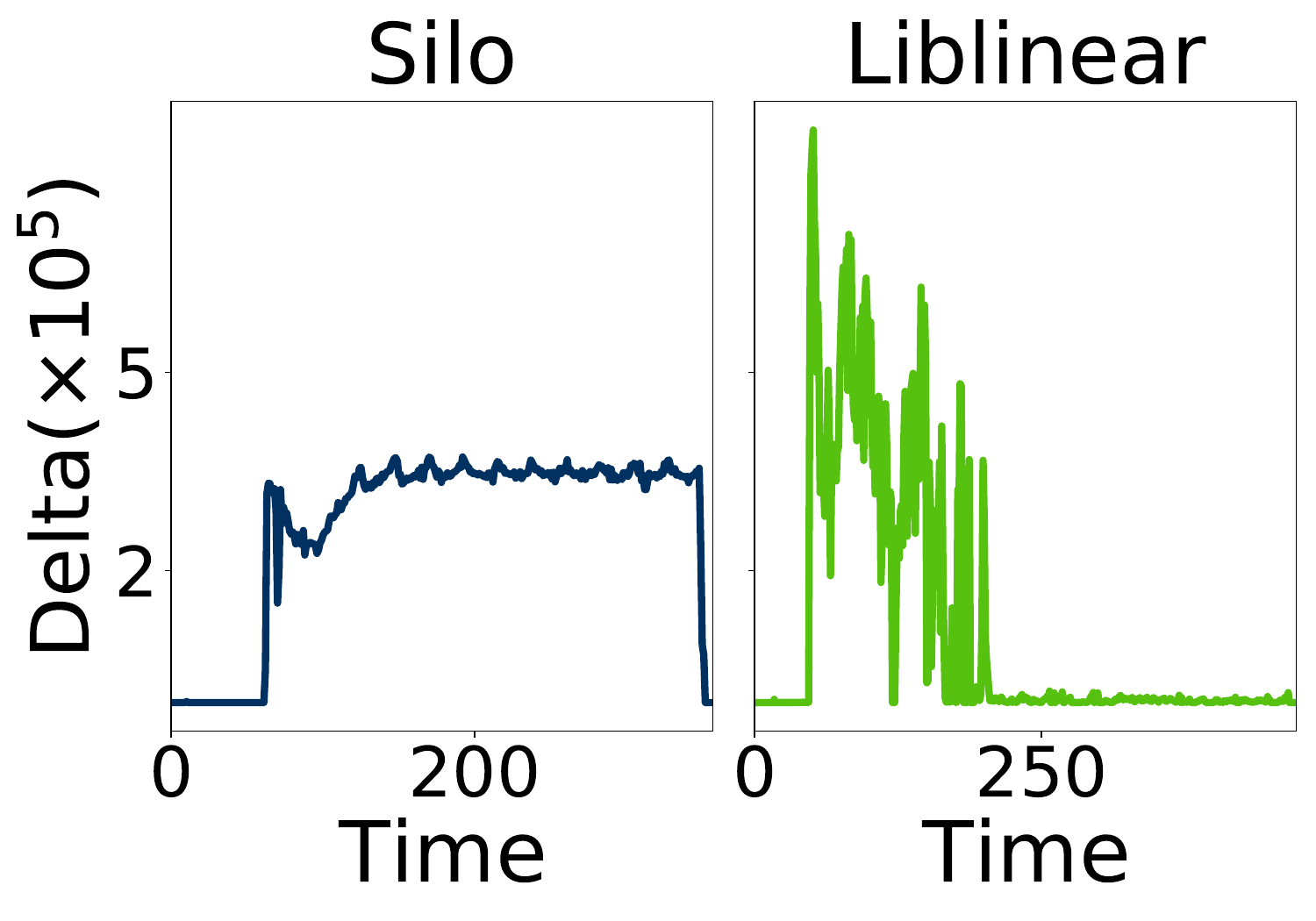}%
        \label{fig:delta}
    }
    \hfill
    \subfloat[Slope]{%
        \includegraphics[width=0.48\linewidth]{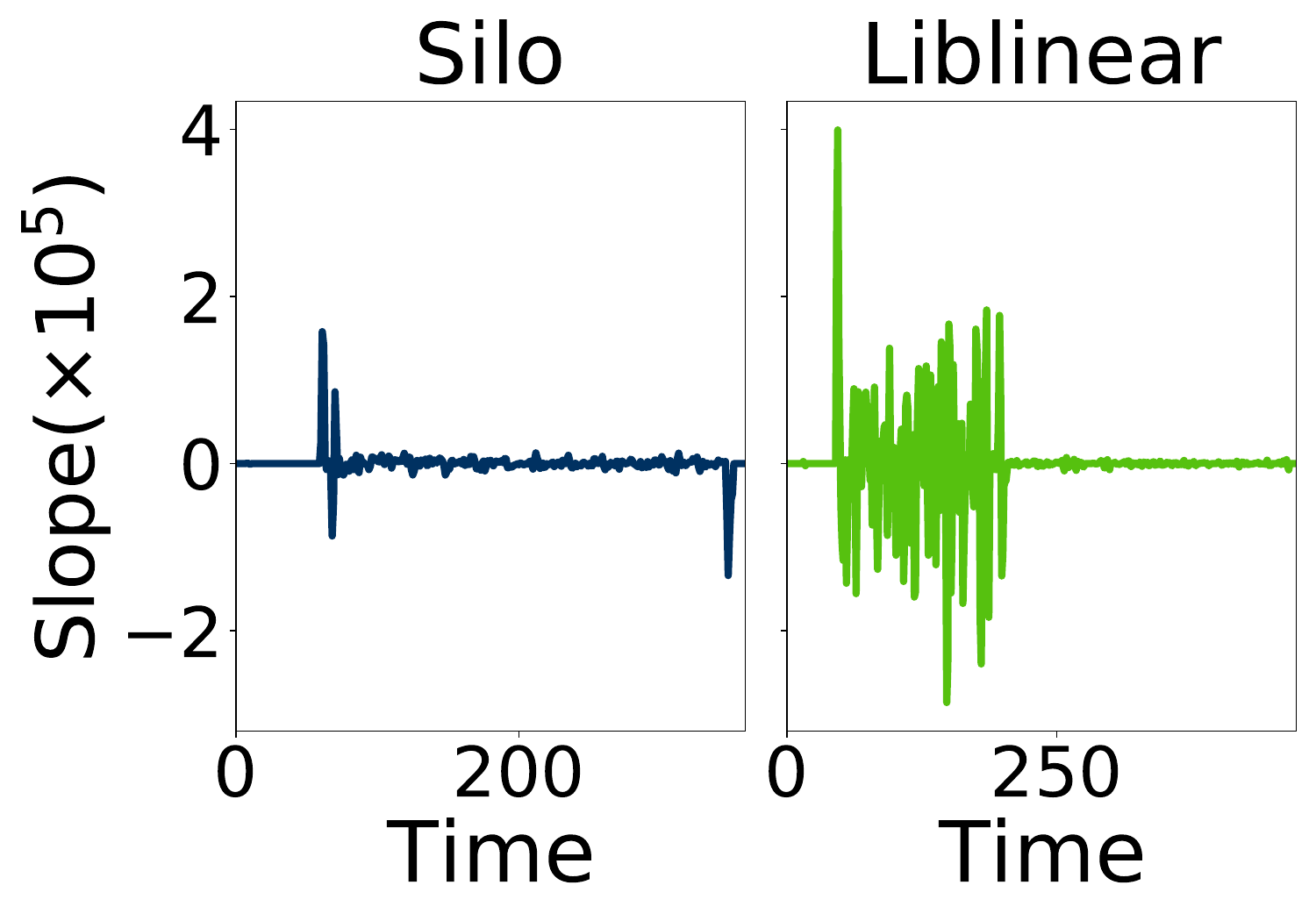}%
        \label{fig:slope}
    }
    \caption{Demote\_promoted delta and its slope of Silo and Liblinear with page migration from TPP-mod.}
        \label{fig:delta_slope}
    \end{figure}

As we mentioned earlier, migration-unfriendly cases can result in repeated page promotion and demotion. In a representative access pattern that leads to a migration-unfriendly one, such as a streaming pattern, pages are promoted to the local DRAM during the first phase but are not accessed afterward. As a result, the utility of DRAM is wasted, while the system continuously promotes pages from the CXL node and demotes pages from the DRAM node. 
In cases where the hot page set is clearly defined but the WSS significantly exceeds the local DRAM size, pages within the hot page set that are relatively less frequently accessed than others will fail to remain in the DRAM. 
These pages will be repeatedly promoted and demoted between DRAM and CXL memory. Conversely, if the hot set can stabilize sufficiently within the local DRAM, the rate at which promoted pages are subsequently demoted becomes minimal, as long as the hot set remains unchanged.
To measure this at runtime, a \texttt{PagePromoted} flag is introduced to track whether a page has been promoted. Additionally, the number of times a page with this flag set is demoted is counted within default Linux's migration step and recorded in \texttt{vmstat} under the metric \texttt{demote\_promoted}. Figure \ref{fig:delta} illustrates the results of measuring the \texttt{demote\_promoted}'s delta at regular intervals during the execution of two workloads: Liblinear with 15 threads and silo with a single thread on a tiered memory system.

\[
    t ={} \text{time};~~p ={} \text{interval for delta}
\]
\[
    \text{delta}(t) ={} \text{demote\_promoted}(t) - \text{demote\_promoted}(t-p)
\]

In the case of Silo, which is the migration-unfriendly workload, the demote\_promoted delta remains consistently high throughout the execution. For the Liblinear, migration-friendly workload, the demote\_promoted delta stabilizes after a certain point in time. If we can terminate migration at the point where the demote\_promoted delta begins to sustain a high level in migration-unfriendly workloads, it is possible to prevent performance degradation caused by excessive migration activity. For migration-friendly workloads, the point where the demote\_promoted delta stabilizes can be considered the stage where the hot set has fully migrated into DRAM. Terminating migration at this stage can reduce overhead caused by excessive fault handling. To establish such criteria on the code level, we compute the slope of the continuously measured demote\_promoted delta using the central difference method. This slope allows for the detection of key points in the workload's execution.

\begin{equation*}
\text{slope}(t) = \frac{\displaystyle 
 (\text{delta}(t) - \text{delta}(t-2p))}{\displaystyle 2}
\end{equation*}
\\
These results are illustrated in Figure \ref{fig:slope}. The point where the absolute value of the slope stabilizes near zero can capture both the onset of a sustained high level of demote\_promoted delta and the point where it stabilizes near zero. By halting migration at these points, it is possible to reduce the overhead caused by migration in migration-unfriendly workloads and achieve optimal performance in migration-friendly workloads by stopping migration after the hot set has stabilized on DRAM. The algorithm for evaluating the aforementioned state is described in Algorithm \ref{alg:earlystop}.
\begin{algorithm}[t]
    \small
    \caption{Earlystop of migration}
    \label{alg:earlystop}

    \KwIn{%
      $Slope_{\text{curr}}$: current demote\_promoted delta's slope\\
      $Slope_{\text{prev}}$: previous demote\_promoted delta's slope\\
      $MaxSlope$: maximum observed slope
    }
    \BlankLine
    \If{$MaxSlope < Slope_{\text{curr}}$}{%
      $MaxSlope = Slope_{\text{curr}}$\\
      $threshold = MaxSlope \gg 2$\\
    }
    \BlankLine
    \If{$SlopeStatement = \text{Varying}$}{%
      \If{$Slope_{\text{prev}} < threshold$}{%
        \If{$Slope_{\text{curr}} < threshold$}{%
          \texttt{allocation is ongoing}\\
        }
        \Else{%
          \texttt{hot page set movement started}\\
        }
      }
      \Else{%
        \If{$Slope_{\text{curr}} > threshold$}{%
          \texttt{hot page set movement ongoing}\\
        }
        \Else{%
          \texttt{hot set movement maybe done} or \\
          \texttt{continuous useless migration started}\\
          $SlopeStatement = \text{Stabilizing}$\\
        }
      }
    }
    \ElseIf{$SlopeStatement = \text{Stabilizing}$}{%
      \If{$Slope_{\text{curr}} > threshold$}{%
        \texttt{hot set should be moved more}\\
        $SlopeStatement = \text{Varying}$\\
      }
      \Else{%
        \texttt{hot set placed on DRAM well} or \\
        \texttt{continuous migration is happening now}\\
        $SlopeStatement = \text{Stabilized}$\\
        % Stop migration indicator
      }
    }
  \end{algorithm}

We set the interval for delta to 2s, thereby the Algorithm \ref{alg:earlystop} executed repeatedly at 2s by a background kernel thread named 'kevaluated' The evaluation of the ping-pong situation is performed based on the recently calculated slope and the previous slope. The slope state is categorized into two states: \texttt{Varying} and \texttt{Stabilizing}. A state where the slope continuously exceeds the threshold is considered \texttt{Varying}, while the opposite condition is considered \texttt{Stabilizing}. The threshold for determining the stabilization of the slope is set proportionally to the maximum slope value. After a slight period of sustained \texttt{Varying} status to confirm enough page movement, the slope state transitions to \texttt{Stabilizing} when a slope below the threshold is measured. Based on this, if the \texttt{Stabilizing} state persists for a certain period, migration will be disabled.

\subsection{Migration Restart}
If changes occur in the dedicated working set selected prior to the termination of migration, it is reasonable to restart migration to ensure that the optimal hot set can be relocated to DRAM. To measure such changes, a page table scan is utilized. After the migration is terminated, a page table scan is performed to measure the number of PTEs (Page Table Entries) with their access bit set by the background kernel thread named krestartd. To avoid excessive overhead from TLB shootdowns which come from clearing access bit, krestartd wakes up every 5 seconds and scans the memory regions corresponding to the entire VM area with the \texttt{VM\_READ} flag set, using a stride of 2MB. We confirmed that using a 2MB stride is sufficient to detect variations in the hot set based on the average of the accessed PTE count. These accessed PTE counts are continuously calculated and maintained in a sliding window to compute an average. If the number of newly calculated counts in the next interval deviates significantly from the average, migration is restarted. The algorithm for each iteration is described in Algorithm \ref{alg:restart}.

\begin{algorithm}[t]
    \small
    \caption{Restart of migration}
    \label{alg:restart}

    \KwIn{%
      $Count_{\text{accessed}}$: current count of accessed PTEs\\
      $Window_{\text{counts}}$: sliding window of past counts\\
      $VariationStatement$: either Varying or Stabilized\\
      $Count_{\text{variation}}$: counter for variation checks
    }
    \BlankLine
    % Compute mean
    $Mean_{\text{accessed}} = \mathrm{mean}(Window_{\text{counts}})$\\
    \BlankLine
    \If{$VariationStatement = \text{Varying}$}{%
      \If{$\lvert Count_{\text{accessed}} - Mean_{\text{accessed}}\rvert < (Mean_{\text{accessed}} \gg 4)$}{%
        $VariationStatement = \text{Stabilized}$\\
      }
      % update window
    }
    \ElseIf{$VariationStatement = \text{Stabilized}$}{%
      \If{$Count_{\text{accessed}} > (Mean_{\text{accessed}} \gg 4)$}{%
        $Count_{\text{variation}} = Count_{\text{variation}} + 1$\\
      }
      \Else{%
        $Count_{\text{variation}} = Count_{\text{variation}} - 1$\\
      }
      % update window
    }
    \BlankLine
    \If{$Count_{\text{variation}} > threshold$}{%
      % restart indicator
      $\Rightarrow$ Restart migration\\
    }
\end{algorithm}Q

The \texttt{VariationStatement}, which indicates the state of hot set variation tracking, is categorized into two states: Varying and Stabilized. Initially, the state is set to Varying immediately after migration is terminated, indicating a condition where the calculated mean and the newly calculated accessed PTE counts are not maintained at a similar level. Once the mean stabilizes and the newly calculated count approaches a similar value to the \texttt{Mean}, the \texttt{VariationStatement} transitions to \texttt{Stabilized}, signifying that changes in the accessed PTE count can be considered indicative of hot set variation. At the beginning of each iteration, a page table scan is performed to calculate the accessed PTE counts, and the mean value is obtained from the \texttt{Window}, excluding the newly calculated value. Then it adjusts the mean or evaluates hot set variation based on the state of the \texttt{VariationStatement}. In the Varying state, the newly calculated count is added to the \texttt{Window} and the iteration concludes immediately to wait for the leveling of the mean at the next iteration. In the \texttt{Stabilized} state, if the difference between the calculated count and mean does not exceed the threshold, the mean is updated and iteration concludes. Otherwise, the count for deciding migration restart increases, and the mean is maintained to enable continuous tracking at the next iteration. After several iterations, if the count for deciding migration restart reaches the predefined threshold, migration is restarted.

\subsection{Multi-tenancy Support}
%-----------------------------------

To effectively achieve our design goals, it is reasonable to enable the decision of whether to perform migration or not at the process level. To enable independent migration toggling at the process level, we added variables to the per task data structure (\texttt{struct task\_struct}) to store the data required for holding the data about demote\_promoted delta's slope and accessed PTE counts which are used for Algorithm ~\ref{alg:earlystop} and ~\ref{alg:restart}. To implement this functionality, we check the owner process of the page and manage the demote\_promoted data of the corresponding process at every kernel level page migration. Also, we add one boolean variable to \texttt{struct task\_struct} to represent the current migration status. \texttt{kevaluated} evaluates the migration friendliness of user-space processes, which is recorded as migration activated, using the demote\_promoted metric described above. \texttt{krestartd} operates in the same manner as \texttt{kevaluated}, except for the condition of a boolean variable, which is the opposite. Through this implementation, we can determine whether to perform migration on a per-process unit using independent data specific to each process. 

From the perspective of the hintfault mechanism, if the page migration of a process has stopped, the poisoning of PTE for hintfault triggering using the \texttt{PTE\_PROTNONE} bit must also cease. Originally, this poisoning task is periodically scheduled and do their process on the currently active process at the scheduled core. Therefore, we modified it to bypass the poisoning task by the status of the process's boolean variable which indicates migration status. Lastly, for the page that remains poisoned even after the migration has been terminated, the critical migration process is just skipped by the same boolean variable.

\subsection{Migration Optimization} \label{migrationopt}
\begin{figure}[t!]
    \centering
    \includegraphics[width=0.48\textwidth]{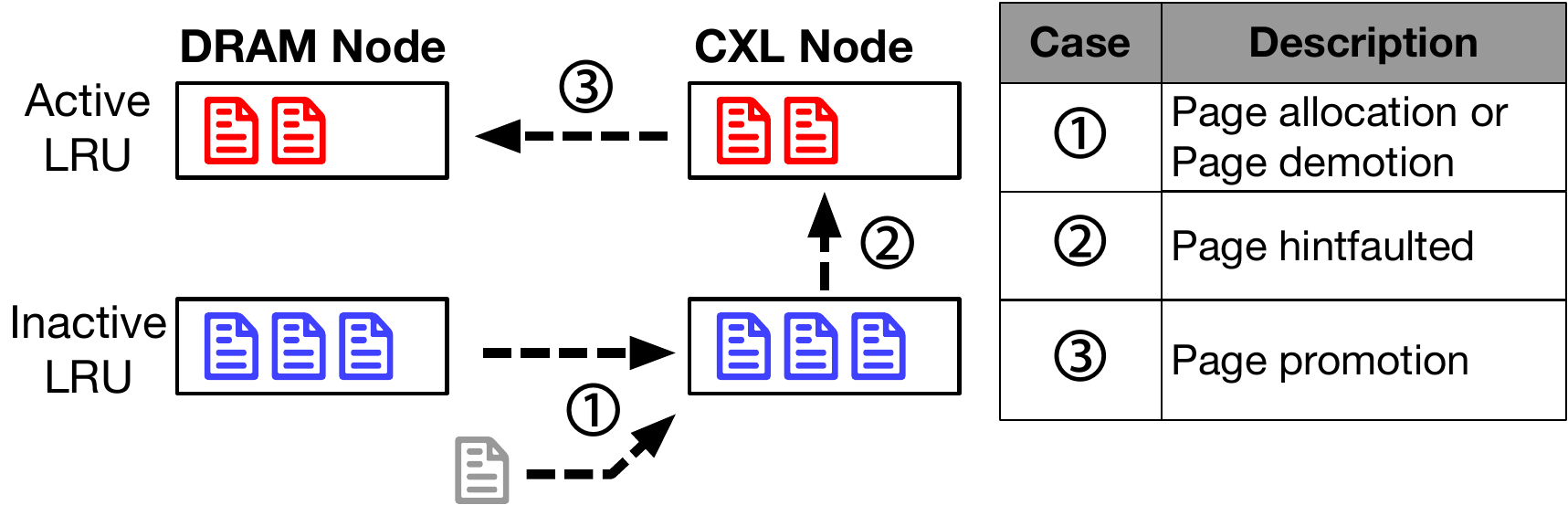}
    \caption{Description of CXL node's LRU aging cases.}
    \label{fig:lruvecagingfigure}
    \vspace{-0.2in}
\end{figure}
%-----------------------------------
%(1)Hintfault 선택 이유 - generality + PEBS를 kernel level에서 쓸때의 불안정성 :: 항상 sampling을 켜두는건 불합리 + 필요할때마다 켤 경우 정상종료가 아닐시 kernel panic\\
\textbf{Profiling mechanism: }
The reason for using TPP as a baseline, specifically hintfault as a profiling mechanism, is to maintain generality and safety. As we know, PEBS can only be used on Intel's CPU. we choose hintfault as a profiling mechanism and TPP as a baseline for generality and safety although there is a similar feature with PEBS on AMD CPUs~\cite{drongowski2010incorporating}. 

\noindent
\textbf{Modified Second Chance LRU: }
TPP maintains page hotness information by using Linux's default LRU page lists. If hintfault occurs on the active list's page, it promotes the page to the upper tier memory node. And, if hintfault occurs on the inactive list's page, it moves the page to the active list. At this point, TPP uses \texttt{mark\_page\_accessed} to move the inactive list's page. When hintfault occurs on the inactive list's page, we expect the page to be moved to the active list instantly, but unfortunately, it is not. When we call \texttt{mark\_page\_accessed}, it adds the page to per CPU's \texttt{pagevec} first. To move the page into the active list, it should wait for batch movement until pagevec is full with pages (in our system, 15 pages are needed). Therefore, the page is still on the inactive list for that period. If hintfault occurs on that page, we think it will be promoted because it's on the active list semantically, but it's not. Hintfault just makes useless excessive fault handling.

To avoid these kinds of problems, we have changed the mechanism for page movement within LRU lists by using the page flag name \texttt{PageHinted}. We use this flag to record hintfault's occurrence on a specific page. The conditions for being a promotion candidate are placing on an active list or \texttt{PageHinted} set. If the active list's page is moved to the inactive list by the absence of recent access, the \texttt{PageHinted} flag will be released. With this small patch, we bypass the excessive occurrence of hintfault while maintaining the same policy with TPP.

\noindent
\textbf{Refault Distance based Hot Page Decision :}
As we mentioned before, access count only-based page hotness detection approaches, which are commonly employed in tiered memory systems, have limitations. \textit{Memory tiering} tried to overcome it by adopting time interval information in migration policy, but its static and global threshold can not afford all workloads because those have diverse access patterns.

Therefore, we trying to use page level access locality by introducing the concept of refault distance, which is partially inspired by the concept of access distance. Refault distance refers to the interval between consecutive hintfault for a specific page. To calculate the refault distance for a specific page, we add age to per NUMA node's LRU, which is aged by page movements within LRU lists. In the modified second chance LRU, the setting of PageHinted is also considered as a movement from inactive list to active list, allowing the interval between hintfault occurrences to be determined by LRU age. The specific cases that trigger node's LRU aging include: Page demotion from DRAM to CXL or initial page allocation on the CXL node(Figure ~\ref{fig:lruvecagingfigure}- \circled{1}) movement from the inactive list to the active list including setting the PageHinted flag cause by hintfault (Figure ~\ref{fig:lruvecagingfigure}- \circled{2}), and promotion of a page in the active list(Figure ~\ref{fig:lruvecagingfigure}- \circled{3}). A separate per NUMA node xarray-based data structure is implemented to store the refault distance information for each page, indexed by the page's PFN (Page Frame Number). 

When a page is placed on CXL by demotion or initial allocation, the CXL node's current LRU age and distance which are initialized to 0 are recorded on xarray. If a hint fault occurs on that page, \texttt{PageHinted} flag will be set and its entry will be updated to the current LRU age and the difference between the age which recorded in the entry and the current LRU age as the first distance. If another hintfault occurs, the second distance will be calculated in the same way. Our design decides to promote that page based on these two distances. If the second distance is shorter than the first distance, we consider the access distance for the page to be decreasing, indicating that it is likely to be accessed again in a short time, and therefore promote it. In the opposite case, we update the xarray only and do not promote the page.

%-------------------------------------------------------------------------------
\section{Evaluation}
\label{sec:figs}
%-------------------------------------------------------------------------------
\subsection{Methodology}
%-----------------------------------
%HW description(table)

\begin{table}[h!]
    \adjustbox{max width=\columnwidth}{%
    \begin{tabular}{ cc|c }
    \hline
    \multicolumn{2}{c|}{\textbf{CPU}}                                                                              & Intel 4th Xeon Gold 2.60Hz / 24 cores                    \\ \hline
    \multicolumn{2}{c|}{\textbf{DRAM}}                                                                             & DDR5 16/32/48GB                               \\ \hline
    \multicolumn{2}{c|}{\textbf{CXL Memory}}                                                                       & Samsung CMM-D (MD210) Prototype                      \\ \hline
    \multicolumn{1}{c|}{\multirow{2}{*}{\textbf{\begin{tabular}[c]{@{}c@{}}Latency\\ (read)\end{tabular}}}} & DRAM & 269 cycles                                    \\ \cline{2-3} 
    \multicolumn{1}{c|}{}                                                                                   & CXL  & 615 cycles                                    \\ \hline
    \multicolumn{1}{c|}{\multirow{2}{*}{\textbf{Bandwidth}}}                                                & DRAM & Read : $\sim$256GB/s, Write : $\sim$248.3GB/s   \\ \cline{2-3} 
    \multicolumn{1}{c|}{}                                                                                   & CXL  & Read : $\sim$17.8GB/s, Write : $\sim$15.8GB/s \\ \hline
    \end{tabular}
    }
    \caption{Hardware configurations.}
    \label{tab:HWenvrionment}
\end{table}

The experiment is conducted on a system configured as a tiered memory system using a real-world CXL memory device which is described in Table \ref{tab:HWenvrionment}. To observe the impact of varying DRAM capacity, memory offlining is used to adjust DRAM capacity to 16GB, 32GB, and 48GB, while the CXL memory is fixed at 128GB. The comparison groups included MEMTIS, NOMAD, TPP-modified (TPP-mod), our work, and a no-migration setup. In our experiment, TPP showed excessive overhead caused by mark\_page\_accessed, making it unsuitable for a proper comparison. As a result, TPP-mod in this study refers to a kernel modified to apply only the modified second chance LRU algorithm on basic TPP. Our work, the no-migration setup, and TPP-mod are implemented on Linux version 5.15, while MEMTIS and NOMAD used publicly available code.  MEMTIS is known to be highly sensitive to the CPU affinity of its two types of kernel threads: one for performing migration and the other for access sampling. Therefore, experiments on MEMTIS are conducted under two configurations. MEMTIS: background kernel threads run on the same cores as the application threads and MEMTIS(+2 core): background kernel threads are pinned to specific CPU cores in a remote socket's CPU only used for them. 

\begin{table}[t]
    \centering
    \footnotesize
    \begin{tabular}{C{2.7cm}|C{2.3cm}}
      \hline
      \textbf{Name} & \textbf{RSS} \\ \hline\hline
      PageRank      & 70.6 GB       \\ \hline
      Silo          & 79.5 GB       \\ \hline
      Liblinear     & 69 GB         \\ \hline
      FT            & 80.1 GB       \\ \hline
      LU            & 92.5 GB       \\ \hline
      SP            & 84.1 GB       \\ \hline
      GUPS (rand)   & 64 GB         \\ \hline
    \end{tabular}
    \caption{Benchmarks for single-tenant evaluation.}
    \label  {tab:single-tenant}
\end{table}

\begin{table}[t]
    \centering
    {\footnotesize
    \renewcommand{\arraystretch}{1.0} % Adjust row height if needed
    \begin{tabular}{C{1.2cm}|C{1.7cm}|C{1.7cm}|C{1.3cm}}
    \hline
    \textbf{Case}       & \textbf{Name} & \textbf{RSS} & \textbf{Threads} \\ \hline
    \multirow{2}{*}{FF} & \textbf{Liblinear}     & 69GB         & 15               \\ \cline{2-4} 
                        & FT            & 40GB         & 24               \\ \hline
    \multirow{2}{*}{FF} & \textbf{Liblinear}     & 72GB         & 15               \\ \cline{2-4} 
                        & SP            & 57GB         & 9                \\ \hline
    \multirow{2}{*}{UF} & \textbf{Silo}          & 53GB         & 1    \\ \cline{2-4} 
                        & FT            & 80.1GB       & 24     \\ \hline
    \multirow{2}{*}{UF} & \textbf{GUPS}          & 64GB         & 12  \\ \cline{2-4} 
                        & SP            & 57GB         & 12              \\ \hline
    \multirow{2}{*}{UU} & \textbf{Silo}          & 60GB         & 1 \\ \cline{2-4} 
                        & GUPS          & 64GB         & 24          \\ \hline
    \multirow{2}{*}{UU} & \textbf{PR}            & 70GB         & 12               \\ \cline{2-4} 
                        & GUPS          & 64GB         & 12               \\ \hline
    \end{tabular}
    }
    \caption{Benchmarks for multi-tenant evaluation.}
    \label{tab:multi-tenant}
\end{table}
The evaluation is conducted with three primary objectives. \textbf{Microbenchmark:} Ensure that our work's behavior conforms to the intended design with microbenchmark which has to change hotset over time. \textbf{Single Tenant:} Compare the execution time of our work (with the complete design applied) against no migration, TPP-mod, MEMTIS, and NOMAD. We divided the scenario into single-threaded(Figure~\ref{fig:eval_2_singleThread}) and multi-threaded(Figure~\ref{fig:eval_2}) execution. \textbf{Multi Tenant:} Compare our work against no migration, TPP-mod, and NOMAD. MEMTIS is excluded from this comparison due to implementation issues, specifically the difficulty in verifying whether two different processes are sharing a single histogram. Multi-tenant experiments are conducted under three cases: FF, UF, and UU. FF indicates the simultaneous execution of two migration-friendly workloads, UU refers to the two migration-unfriendly workloads, and UF represents two workloads with differing migration-friendliness. In the multi-tenant, two applications will be initiated with varying starting time offsets to ensure the scenario that the two benchmarks occupy different proportions of DRAM at their initial allocation. In each pairing, the application that is launched first is highlighted in bold in Table~\ref{tab:multi-tenant}. The second application is subsequently initiated after a predetermined time offset relative to the start time of the first application. The single-tenant experiments are conducted on each DRAM size (16GB, 32GB, and 48GB), and the multi-tenant experiments and Refault distance experiments are conducted with the DRAM size fixed at 32GB. A specific description of the used benchmarks and RSS of each are listed in Table \ref{tab:single-tenant} and \ref{tab:multi-tenant}.

\subsection{Microbenchmark}
\begin{figure}[t]
    \centering
    {\includegraphics[width=1.0\linewidth]{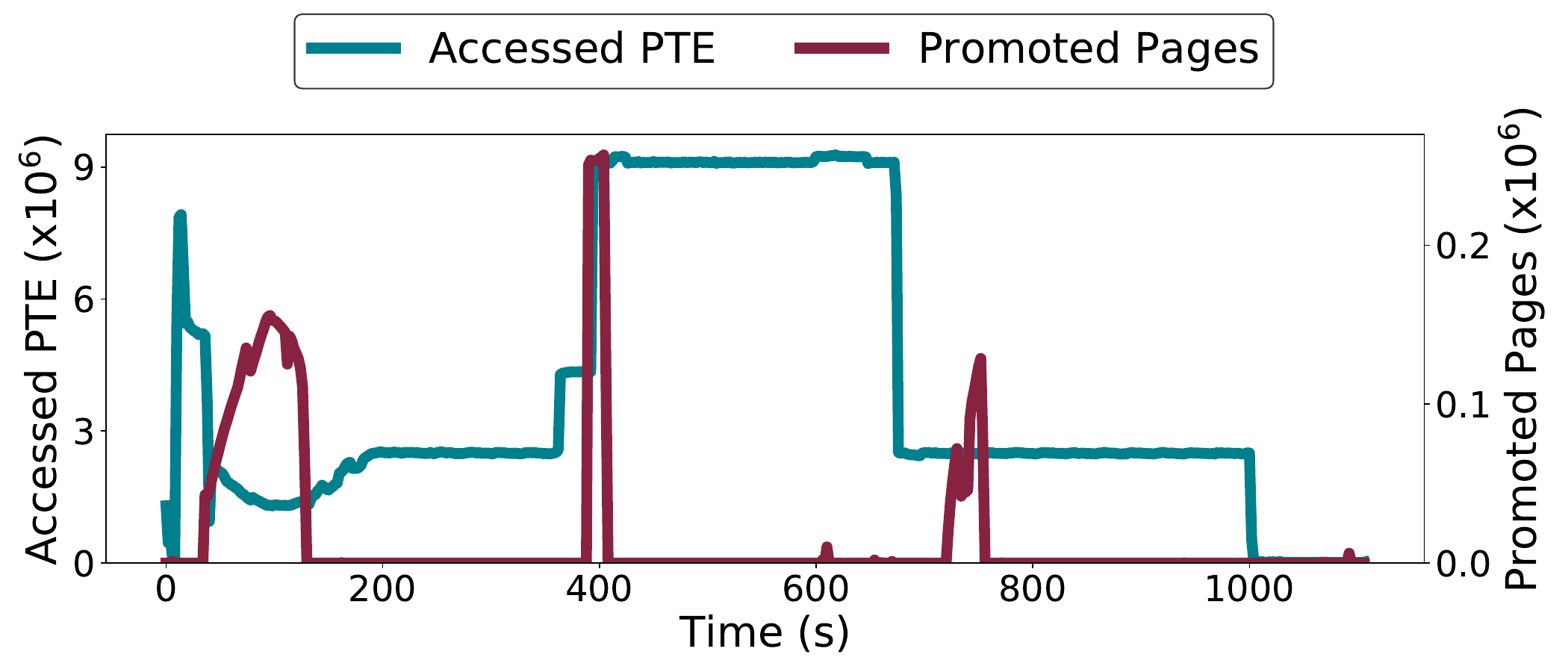}}

    \vspace{-0.5em}
% \vspace{-0.2in}
    \caption{Our Work's behavior on microbenchmark. 'Accessed PTE' shows the number of PTE whose access bit set over time. 'Promoted pages' show the amount of page promotion performed by our work.}
    \vspace{-0.1in}
        \label{fig:microbench}
\end{figure}
% \begin{microbenchmark-figure}
% \end{microbenchmark-figure}
To ensure that our work's behavior conforms to the design, we conduct a experiment with a microbenchmark. Microbenchmark uses 80GB of RSS and makes dedicated access to randomly selected 30GB of pages(first phase), then loosens the access range to 60GB while changing the access pattern(second phase). Finally, intensive access is performed again on the initially selected 30GB region(third phase). Each step lasts for 5 minutes. From an omniscient perspective, migration should be stopped well in each phase due to the mismatch in WSS and restarted at the transition points between phases. 

Figure~\ref{fig:microbench} shows the number of page promotions in our work and the results of the page table scan performed on the entire memory region used by the microbenchmark over time. The memory configuration is set to 16GB:128GB. As shown in the graph, our work triggers three migration stops and two restarts during the lifecycle of the microbenchmark which is equal to the best option. In the first phase, our work continuously attempts to promote the 30GB hot set to DRAM. Over time, it recognized the migration-unfriendly situation caused by the mismatch between the WSS and DRAM and subsequently stopped migration. At the beginning of the second phase, our work detects a variation in the hot set through a page table scan and restarts migration accordingly. After the restart, migration is stopped again in a relatively shorter period compared to the first phase. This can be attributed to the 60GB region selected during the second phase already includes most of the hot set from the first phase which is already residing in DRAM. This allowed for quicker detection of changes in the demote\_promoted metric. Finally, in the third phase, the system again performed a similar cycle of restart and stop. These observations confirm that our work effectively handles not only migration stops based on migration-friendliness but also hot set variation detection through page table scan.

\subsection{Single-tenancy Results}
%-----------------------------------
\begin{figure*}[!t]
    \centering
    \begin{minipage}[t]{0.49\linewidth} 
        \centering
        \includegraphics[width=\linewidth]{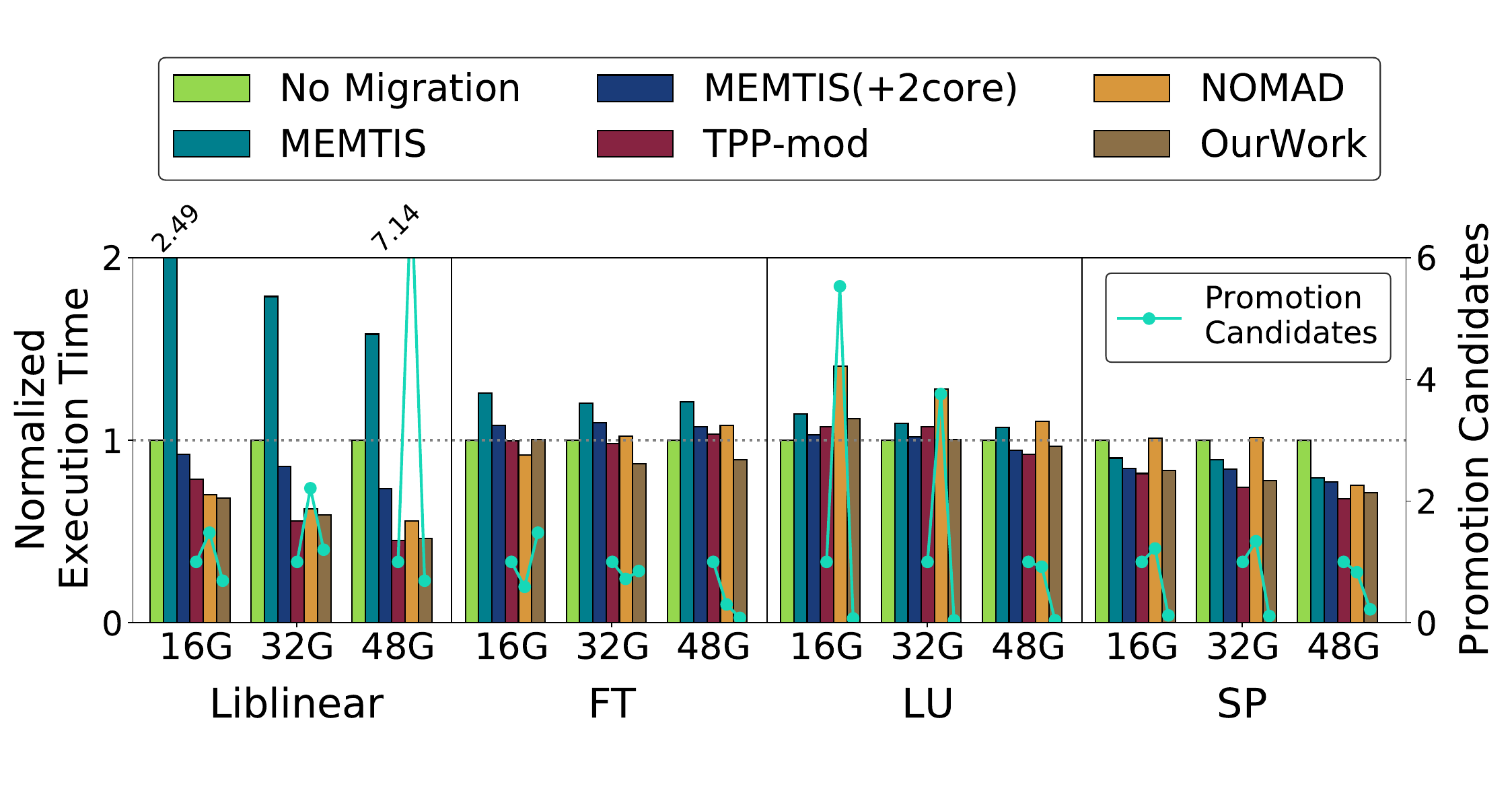}
        \vspace{-0.4in}
        \caption*{(a) Comparisons for Migration-friendly benchmarks. } 
    \end{minipage}
    \hfill 
    \begin{minipage}[t]{0.49\linewidth} 
        \centering
        \includegraphics[width=\linewidth]{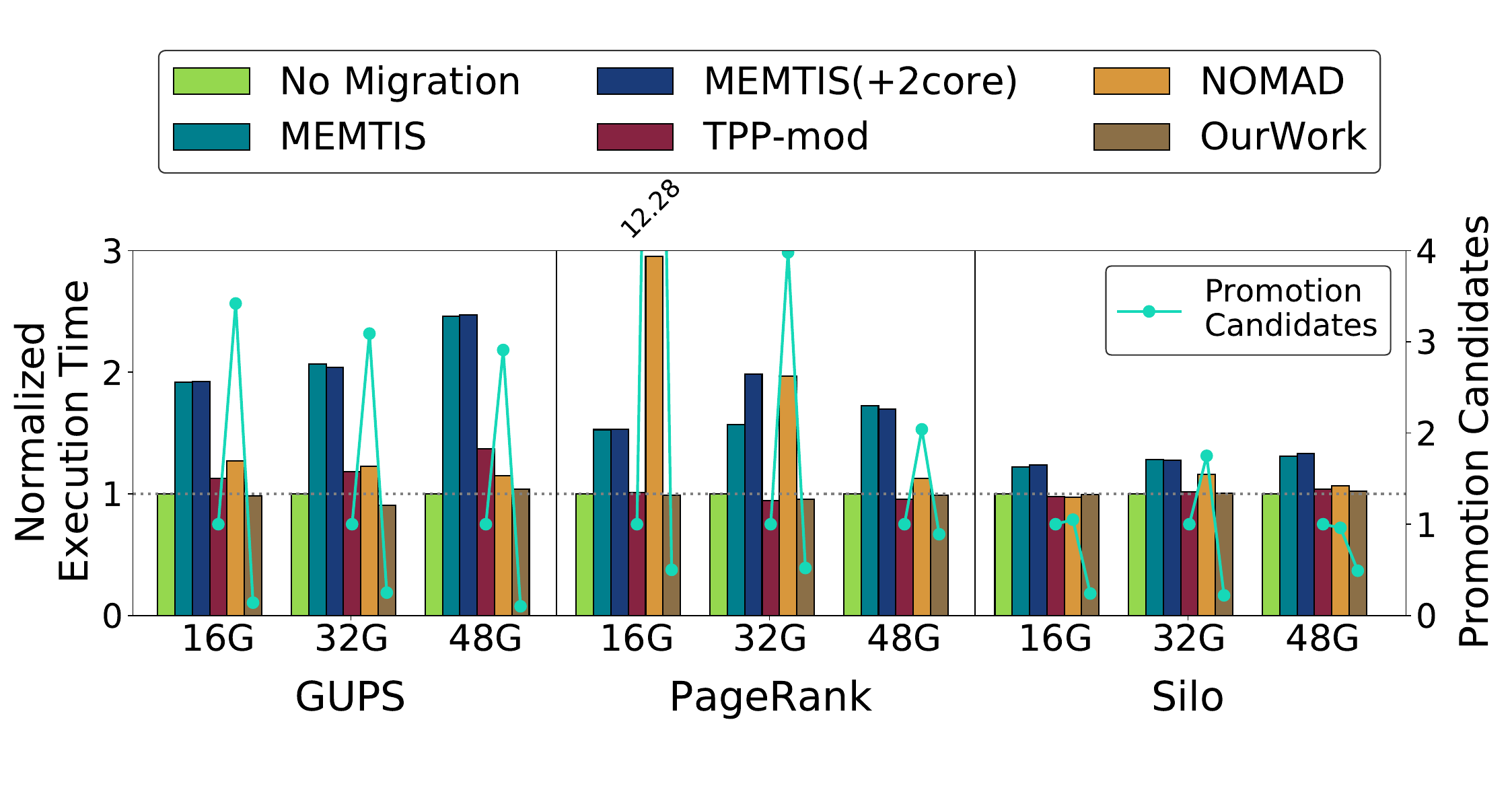}
        \vspace{-0.4in}
        \caption*{(b) Comparisons for Migration-unfriendly benchmarks.} 
    \end{minipage}
    \vspace{-0.1in}
    \caption{Evaluation comparison for single-tenant multi-threaded benchmarks. Performances are normalized to 'No migration'. The capacity listed represents configured DRAM capacity.  The number of trial for page promotion is represented as 'Promotion Candidates' and applies to NOMAD, TPP-mod, and OurWork. }
        \label{fig:eval_2}
\end{figure*}

\begin{figure}[!t]
    \centering
    {\includegraphics[width=1.0\linewidth]{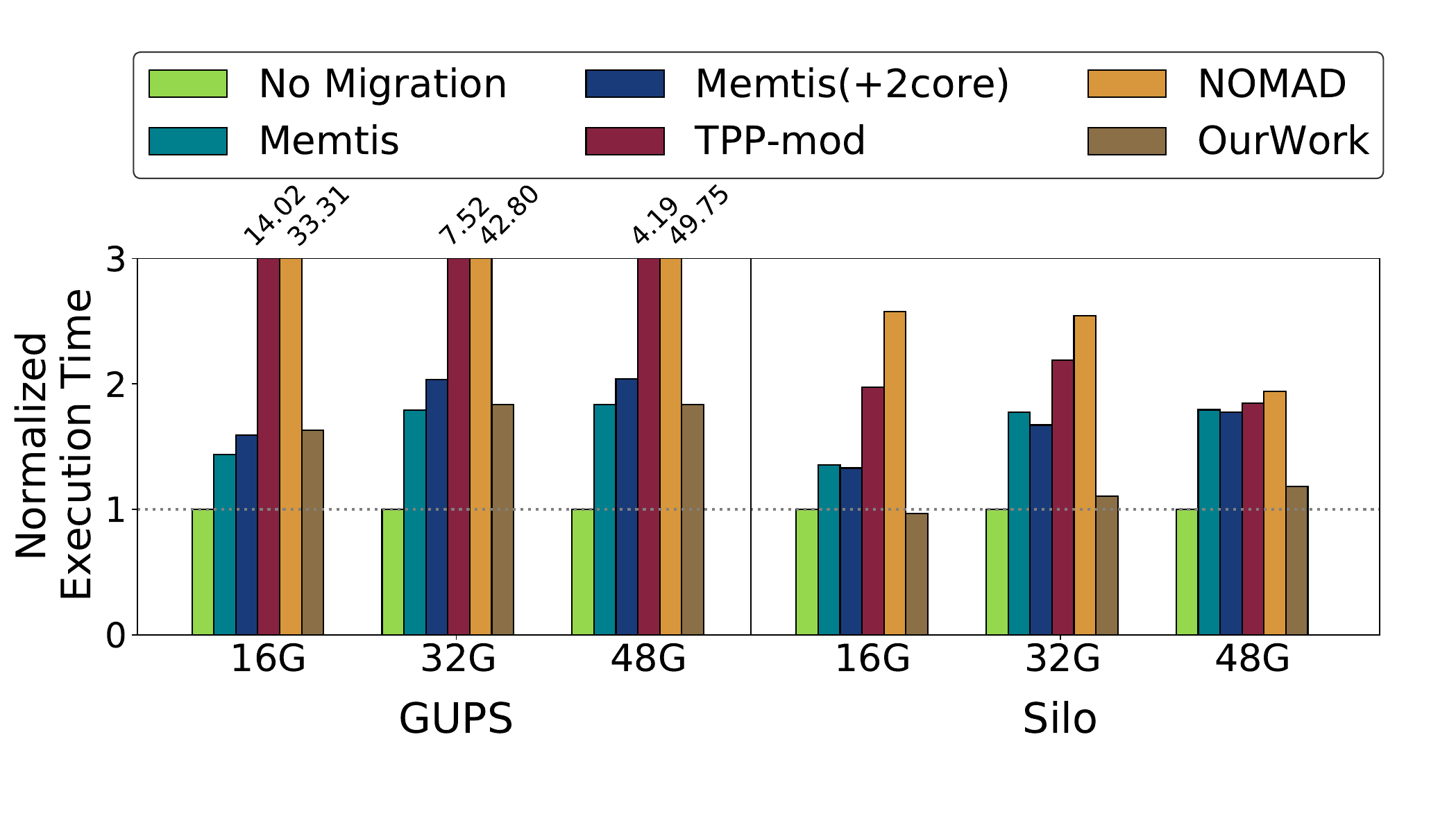}}

    \vspace{-0.2in}
% \vspace{-0.2in}
    \caption{Evaluation for normalized execution time on single-threaded benchmarks.}
    \vspace{-0.1in}
        \label{fig:eval_2_singleThread}
\end{figure}

Figure~\ref{fig:eval_2} presents the execution times of the benchmarks. In migration friendly benchmarks, hotness-based migration schemes, including the our work proposed in this study, exhibit superior performance compared to schemes that do not utilize migration. But, migration schemes other than ours tend to excessively migrate pages even under migration unfriendly workloads, thereby negatively impacting application performance.

In Liblinear benchmark, every scheme excluding MEMTIS shows better performance than no migartion scenario which indicates that libnear is migration friendly benchmark. The proposed scheme shows 54\% better performance than no migration in 48GB of dram. However, FT's performance shows different results. Proposed schemes outperform no migration different from others. Ours shows a performance improvement of 13\% compared to no migration. Nevertheless, MEMTIS has about 20\% performance degradation. This result is due to the loss of temporal information when choosing the hot set.

NOMAD exhibits inferior performance compared to TPP-mod and our proposed scheme in the PageRank benchmark especially in 16GB size of DRAM which exhibits 13x lower performance than no migration. Although NOMAD has adopted asynchronous migration to accelerate the migration process, the working set size for PageRank exceeds 32GB. Consequently, with DRAM sizes of 16GB and 32GB, the excessive migration overhead of NOMAD degrades performance, as illustrated in the graphs. 
% 부가 설명, migration firend할 때는 nomad가 괜찮을 수 있음.
When the DRAM size is approximately 48GB, which is smaller than the PageRank working set size, NOMAD demonstrates performance comparable to TPP-mod, as expected.

In the single-threaded GUPS benchmark, which is migration unfriendly shown in Figure~\ref{fig:eval_2_singleThread}, histogram based hot set selection method such as MEMTIS exhibits better performance than other schemes. The number of migrations in MEMTIS is smaller than in other schemes because the histogram is not fully utilized in a single-threaded environment. Paradoxically, in this context, MEMTIS cannot accurately capture the hot set, and therefore it outperforms other migration schemes.

Silo exhibits negligible effects from page migration in a multi-threaded environment, demonstrating its neutral migration-friendliness in multi-thread.
However, in a single-threaded environment, migration-unfriendliness increases, leading to execution time growth rates of 1.32–2.54x for other comparison works across all memory configurations. In contrast, our work limits the maximum execution time growth to only 1.17x.

\begin{figure*}[!t]
    \centering
    \begin{minipage}[t]{0.49\linewidth}
        \centering
        \includegraphics[width=\linewidth]{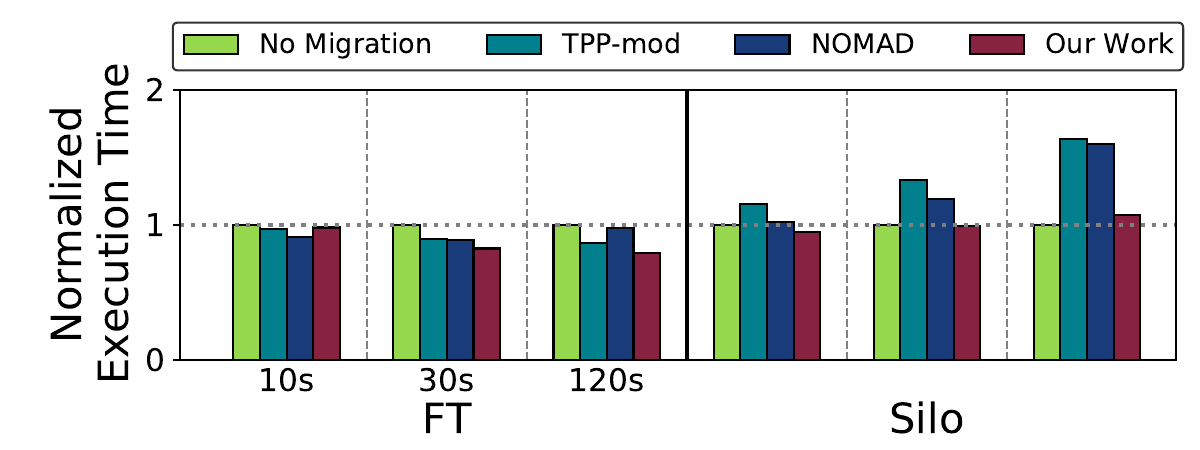}
        \vspace{-0.3in}
        \caption*{(a) Results with varying start time offset about UF case.} 
    \end{minipage}
    \hfill 
    \begin{minipage}[t]{0.49\linewidth}
        \centering
        \includegraphics[width=\linewidth]{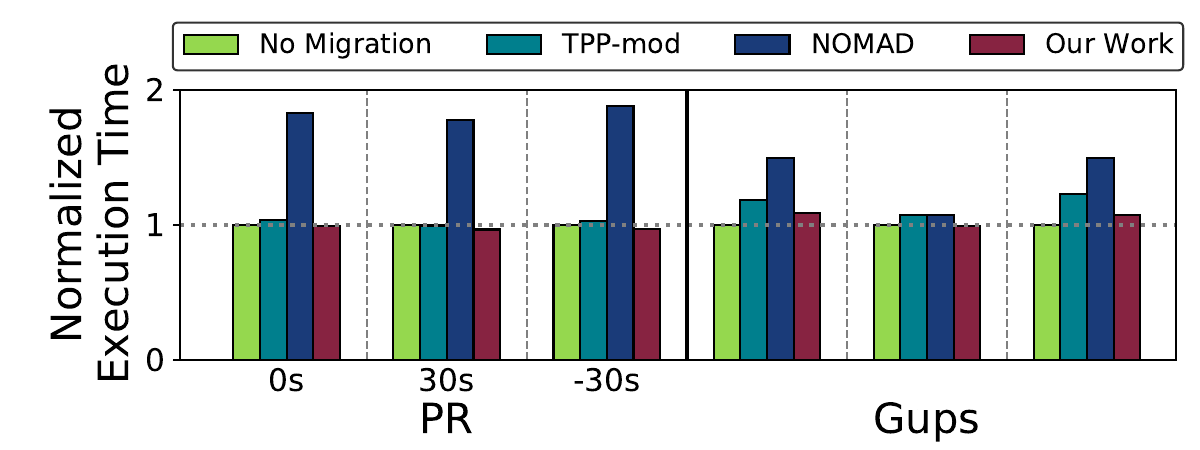}
        \vspace{-0.3in}
        \caption*{(b) Results with varying start time offset about UU case.}
    \end{minipage}
    \vspace{-0.1in}
    \caption{Evaluation results of execution time on multi-tenant environment. The start time offset is pre-determined to vary the DRAM occupancy of the two workloads at their initial allocation. The benchmark without the offset listed above its name starts first. Comparison groups' values are normalized to 'No Migration' on each benchmark-time offset. }
        \label{fig:eval_3_pick}
\end{figure*}

\begin{figure*}[!t]
    \centering
    \begin{minipage}[t]{0.33\linewidth} 
        \centering
        \includegraphics[width=\linewidth]{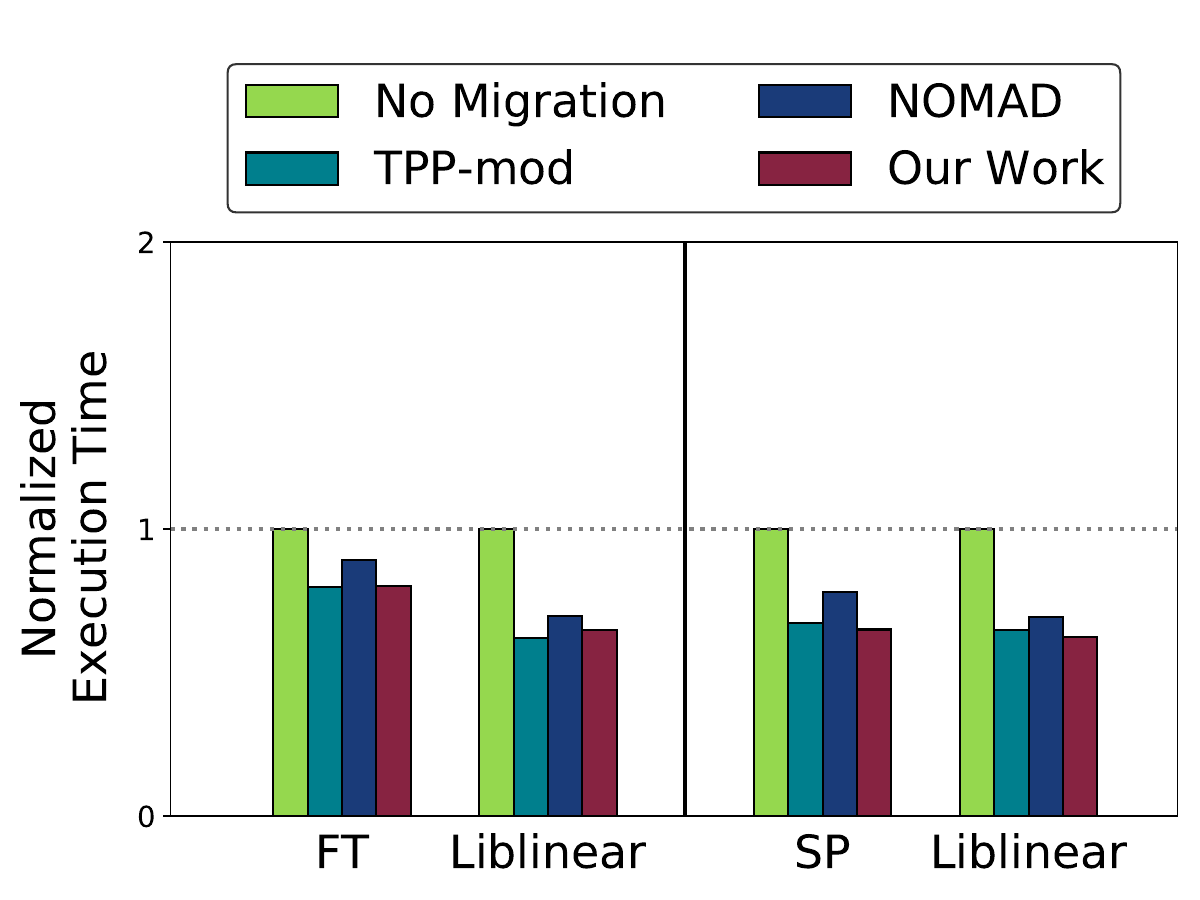}
        \vspace{-0.3in}
        \caption*{(a) FF cases}
    \end{minipage}
    \hfill  % 간격 조정
    \begin{minipage}[t]{0.33\linewidth} 
        \centering
        \includegraphics[width=\linewidth]{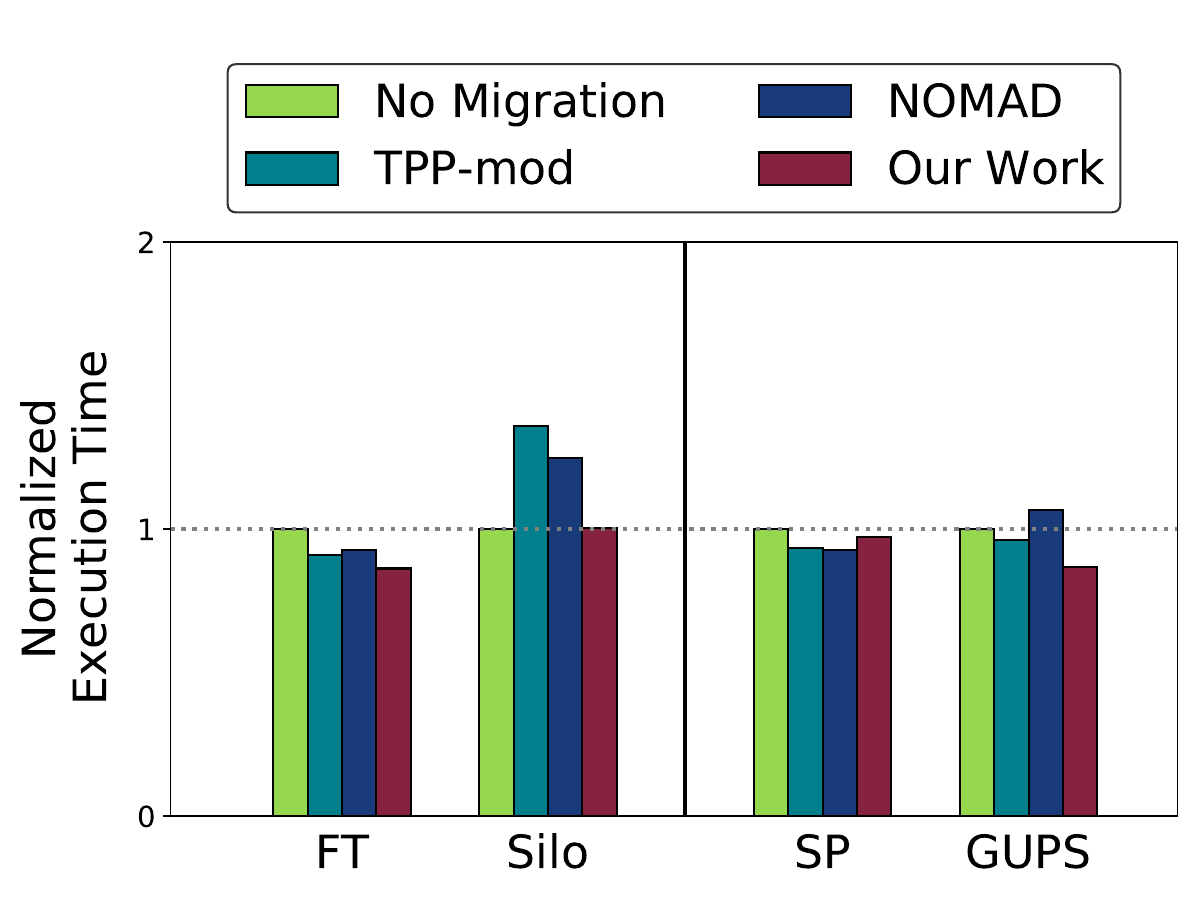}
        \vspace{-0.3in}
        \caption*{(b) UF cases} 
    \end{minipage}
    \hfill  % 간격 조정
    \begin{minipage}[t]{0.33\linewidth} 
        \centering
        \includegraphics[width=\linewidth]{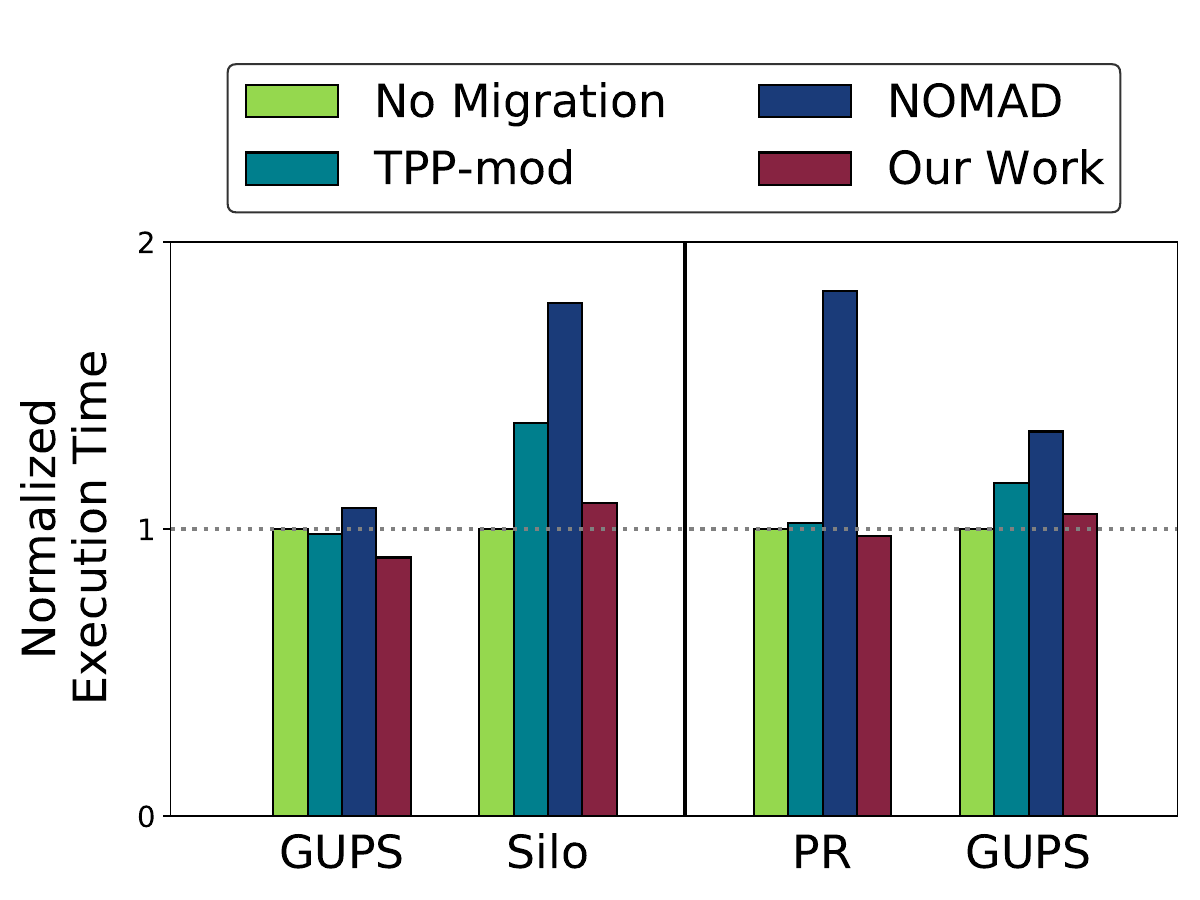}
        \vspace{-0.3in}
        \caption*{(c) UU cases} 
    \end{minipage}
    \caption{Comparison graph of average values for varying start time offsets. Each pair executed for 3 time with different offset. Values are normalized to average of 'No Migration' of each benchmark.}
        \label{fig:eval_3}
\end{figure*}

\subsection{Multi-tenancy Results}
%-----------------------------------

Figure~\ref{fig:eval_3} presents the evaluation results for the multi-tenant scenario. We tested various configurations, including:
1. Migration-friendly benchmark paired with migration-friendly benchmark (FF);
2. Migration-unfriendly benchmark paired with migration-friendly benchmark (UF); 
3. Migration-unfriendly benchmark paired with migration-unfriendly benchmark (UU).

In the FF scenario, all migration schemes demonstrated superior performance compared to the no-migration scheme. Notably, the SP and Liblinear components of our proposed scheme outperformed the no-migration scheme by approximately 35\% and 38\%, respectively, as anticipated.

In scenarios where the characteristics of the benchmarks differ, such as the UF scenario, the access patterns of each benchmark can become diluted. This dilution makes it challenging to assess the overall migration friendliness of the entire system. However, our migration friendliness decision mechanism effectively addresses this issue by implementing per-process migration stops and restarts. This approach remains valid in the UU scenario.

Furthermore, the evaluation results demonstrate that the proposed scheme maintains resilience across varying DRAM usage scenarios. Variations in benchmark start times, as depicted in Figure~\ref{fig:eval_3_pick}, lead to differing DRAM allocations between the two tenants. Under these conditions, alternative approaches exhibit unstable behavior depending on the initial DRAM size, particularly in the case of Nomad, where performance degradation approaches approximately 88\% compared to no migration. However, in almost all instances, our approach delivers enhanced performance compared to existing methods.

In summary, the majority of evaluation outcomes indicate that our proposed solution outperforms other migration strategies across diverse multi-tenant configurations.

%-------------------------------------------------------------------------------
\section{Conclusion}
\label{sec:figs}
%-------------------------------------------------------------------------------
Every workload has its own migration-friendliness based on its access pattern and system configuration. 
We show that the per-page ping-pong situation can be an appropriate metric for evaluating the effectiveness of migration.
Additionally, migration-friendliness-based migration toggling can effectively prevent performance degradation which comes from migration overhead. 
It prevents performance degradation on migration-unfriendly workloads and enhances performance on migration-friendly workloads. 
Also, it can reliably handle multi-tenant environments as well by toggling migration at a per-process level with low cost. 
It effectively turns off migration, providing an average of 14.8\% performance improvement compared to NOMAD with migration-friendly workloads, while providing an average of 36.0\% improvement with migration-unfriendly workloads.
Additionally, our scheme achieves up to 72.0\% performance improvement compared to NOMAD in multi-tenant environments, exhibiting that migration toggling is effectively applied at the per-process level.

%-------------------------------------------------------------------------------
\section*{Availability}
%-------------------------------------------------------------------------------

Our implementation based on Linux v5.15 for whole design incorporated is available. The source code will be open to public after publication.

%-------------------------------------------------------------------------------

\bibliographystyle{plain}
\bibliography{output}

%%%%%%%%%%%%%%%%%%%%%%%%%%%%%%%%%%%%%%%%%%%%%%%%%%%%%%%%%%%%%%%%%%%%%%%%%%%%%%%%
\end{document}